\documentclass[journal,10pt]{IEEEtran}

\usepackage{wrapfig}
\usepackage[usenames, dvipsnames]{color}
\usepackage{graphicx}
\usepackage{pdfpages}
\usepackage{epstopdf}	
\usepackage{amsmath}
\usepackage{mathrsfs}
\usepackage{longtable}
\usepackage{enumerate}
\usepackage{algorithmic}
\usepackage{multirow}
\usepackage{mathtools}
\usepackage{cite}
\newcommand{\ignore}[1]{}
\usepackage{caption}
\usepackage{subcaption}
\usepackage{fix-cm}    
\usepackage{amssymb}
\usepackage{array}
\makeatletter
\newcommand\HUGE{\@setfontsize\Huge{50}{60}}
\makeatother  

\ignore{
\makeatletter
\def\ps@headings{%
\def\@oddhead{\mbox{}\scriptsize\rightmark \hfil \thepage}%
\def\@evenhead{\scriptsize\thepage \hfil \leftmark\mbox{}}%
\def\@oddfoot{}%
\def\@evenfoot{}}
\makeatother
\pagestyle{headings}
}

\renewcommand{\baselinestretch}{1}

\newtheorem{theorem}{Theorem}

\newtheorem{remark}{Remark}

\newcommand{\argmax}{\operatornamewithlimits{argmax}}
\newcommand{\argmin}{\operatornamewithlimits{argmin}}

\begin{document}
\setlength{\belowdisplayskip}{3pt}

\title{\LARGE \bf Stability Analysis of Simple and Online User Association Policies for Millimeter Wave Networks}

\author{Vaibhav Kumar Gupta, Santosh Kumar Singh, and Gaurav S. Kasbekar}

\IEEEoverridecommandlockouts

\maketitle

{\renewcommand{\thefootnote}{} \footnotetext{V. K. Gupta, S. K. Singh and G. Kasbekar are with Department of Electrical
Engineering, Indian Institute of Technology (IIT), Bombay. Their email
addresses are \{vaibhavgupta, santoshiitb, gskasbekar\}@ee.iitb.ac.in, respectively. \\

}}
\begin{abstract}
In a millimeter wave (mmWave) network, user association-- the process of deciding as to which base station (BS) a given user should associate with-- is a crucial process which affects the throughput and delay performance seen by users in the network and the amount of load at each BS. In the existing research literature, the stability region of a user association policy, \emph{i.e.}, the set of user arrival rates for which the user association policy stabilizes the network, has not been analytically characterized for any user association policy for
mmWave networks. In this paper, we study the user association problem in
mmWave networks and compare the performances of four user association policies: Signal to Noise Ratio (SNR) based, Throughput based, Load based and Mixed. All these policies are simple, easy to implement, distributed
and online. We use a Continuous Time Markov Chain (CTMC) model and Lyapunov function techniques to analytically characterize the stability region of each of the above four user association policies. We also evaluate the performances of
the above four user association policies in a large mmWave network, in which link qualities fluctuate with time and users are mobile, via detailed simulations. Our results show that the Throughput based policy outperforms the other three user association policies in terms of stability region as well as average throughput, average
delay and fairness performance.
\end{abstract}
\begin{IEEEkeywords}
mmWave Networks, User Association, Continuous Time Markov Chain, Stability. 
\end{IEEEkeywords}
\section{Introduction}
\label{introduction}
Millimeter wave (mmWave) networking, which is a key fifth generation (5G) cellular network  technology,  is receiving significant attention from the research community due to its potential to effectively support high data rate wireless applications, \emph{e.g.,} large file transfers, high definition video streaming etc~\cite{rangan}. The currently deployed 4G Long Term Evolution Advanced (LTE-A) technology is faced with the spectrum scarcity problem. In contrast, mmWave communication takes place in high frequency bands-- from 30 to 300 GHz-- most of which are currently unused. Utilization of these high frequency bands significantly increases the amount of available bandwidth~\cite{akdeniz}. However, it is challenging to communicate using mmWave bands since they suffer from heavy  attenuation of transmissions due to high path loss, low diffraction and high sensitivity to shadowing and blockage effects. Also, to overcome the high path loss, directional communication is used, which results in rapid variation of the qualities of links between  transmitters and receivers. In addition, due to the large bandwidths used and directivity of transmissions, mmWave networks are noise limited, in contrast to sub-6GHz wireless networks, which are interference limited; this results in new challenges in mmWave networking~\cite{what5g}. 
In cellular networks, irrespective of the technology used, each user needs to find a base station (BS) to associate with before exchanging its data. User association is a crucial process which affects the throughput and delay performance seen by users in the network and the amount of load at each BS~\cite{Hong,ritesh,kyuho}. The user association problem, \emph{i.e.}, the problem of deciding as to which BS a given user should associate with, has been extensively studied in the context of sub-6GHz wireless networks, e.g., in~\cite{Hong,ritesh,kyuho,Han,awais,yigal,gsk,qiao}.  The signal to noise ratio (SNR) based association mechanism has been adopted by the IEEE 802.15.3c and IEEE  802.11ad mmWave network standards~\cite{ocalb}; in this mechanism, each user associates with the BS with which it has the largest SNR. However, the SNR based association mechanism can result in an uneven load distribution in the system, with  a few BSs being heavily loaded and others being underutilized, poor throughput and delay performance achieved by users or even instability of the network~\cite{outlook},~\cite{MDP}. User association schemes for mmWave networks have been proposed in~\cite{ocalb,MDP,energy,cbss,power,sharing,thr,angela,fractional,abram,dacr,maua,duca,uawbb,ua5g}; see Section~\ref{lit} for a review. However, for most user association schemes proposed in the prior literature, there are no proven guarantees on performance. In particular, to the best of our knowledge, in the existing research literature, the stability region of a user association policy, \emph{i.e.}, the set of user arrival rates for which the user association policy stabilizes the network,  has not been analytically characterized for \emph{any} user association policy for mmWave networks. This is the space in which we seek to contribute in this paper. In addition, in the model in most prior works, it is assumed that the number of users in the system is constant. In contrast, in the model in this paper, users arrive into and depart from the system over time, as in practice.

In this paper, we study the user association problem in mmWave networks and compare the performances of the following four user association policies: SNR based, Throughput based,  Load based and Mixed. In the SNR based policy, a user associates with the BS with which it has the highest SNR. In the Throughput based policy, a user associates with the BS with which it obtains the highest throughput.  In the Load based policy, a user associates with the BS with the  smallest number of existing associated users. In the Mixed policy, a user chooses the BS for which a linear combination of throughput and data rate achieved by the user is maximized. All these policies are simple, easy to implement, distributed and online, in contrast to user association algorithms proposed in several prior works, which are centralized and/ or require the system to solve a complex optimization problem. To implement these policies, a user only requires knowledge of the qualities of the links from all the BSs in its range to itself, and the numbers of users that are already associated with each BS in its range. This information can be easily obtained with the help of reference  signals sent from each BS. 
We investigate the stability regions of the above four user association policies using a Continuous Time Markov Chain (CTMC)~\cite{asmussen} model. First, for analytical tractability, we consider a simplified network model with two BSs. Users arrive according to a Poisson process and experience one of two possible data rates with each of the two BSs, depending on whether they have a Line of Sight (LoS) or Non-LoS (NLoS) link with the BS. Each arriving user selects one of the two BSs to associate with, using one of the above four user association policies. We use Lyapunov function techniques~\cite{asmussen}  to analytically characterize the stability region of each of the above four user association policies. Next, we generalize several of the results obtained via stability analysis of the above simplified two BS model to the case where there are an arbitrary number of BSs. Also, we validate the obtained analytical results via simulations. We also evaluate the performances of the above four user association policies in a large mmWave network, in which link qualities fluctuate with time and users are mobile, via detailed simulations. Our results show that the Throughput based policy outperforms the other three user association policies in terms of stability region as well as average throughput, average delay and fairness performance. In particular, the Throughput based policy consistently outperforms the Mixed policy. This is surprising since under a model similar to that in this paper, but in the context of sub-6GHz 802.11 Wireless Local Area Networks (WLANs), the Mixed policy was found to outperform the Throughput based policy~\cite{gsk} (see Section~\ref{lit} for details). In addition, our results show that the performance of the SNR based policy is the worst among all the four user association policies.

The rest of this paper is organized as follows. The related research literature is reviewed in Section~\ref{lit}. Our network and channel models are described in Section~\ref{network model}. The various user association policies whose performance is studied are formally defined in Section~\ref{policy}.  The stability regions of these policies under the two BS model are characterized   in Section~\ref{simplemodel}. In Section~\ref{gen_case}, the results obtained via stability analysis in Section~\ref{simplemodel} are generalized to the case where there are an arbitrary number of BSs. The performances of these policies are evaluated via simulations in Section~\ref{simulation} and conclusions are provided in Section~\ref{conclusion}.  

\section{Related Work}
\label{lit}
The user association problem has been extensively studied in the context of sub-6GHz wireless networks, e.g., in~\cite{Hong,Han,ritesh,kyuho,awais,yigal,gsk,qiao}. 
In particular, an $\alpha-$optimal user association algorithm for 4G cellular networks has been proposed in~\cite{Hong}. In this algorithm, different values of the parameter $\alpha$ correspond to different  association rules-- SNR based, throughput based, delay based and load based.  User association taking into account load-balancing in heterogeneous networks (HetNets)  has been studied as a utility maximization problem in~\cite{qiao}. A distributed algorithm based on dual decomposition has been proposed. The performance of online user association policies for IEEE 802.11 WLANs has been studied and the range of user arrival rates for which the system is stable has been analyzed in~\cite{gsk}. The SNR, Throughput and Mixed policies in this paper are similar to the SNR, Selfish and RAT policies, respectively, in~\cite{gsk}. However, there are several differences between the model in~\cite{gsk} and that in this paper. In particular, in the model in this paper, a BS can have multiple radio frequency (RF) chains~\cite{shokri} and hence, a BS $i$ can simultaneously communicate with up to $m_i$ users. On the other hand, in the model in~\cite{gsk}, each BS communicates with at most one user at a time. Also, in the model in~\cite{gsk}, if $\mathcal{X}_i$ is the set of users associated with BS $i$ and $r_{ij}$ is the data rate of user $j$ associated with BS $i$, then the throughput that each user associated with BS $i$ gets is $\frac{1}{\sum\limits_{j \in \mathcal{X}_i}\frac{1}{r_{ij}}}$, whereas in the model in this paper, the throughput is given by~\eqref{thr}. The Load based policy, whose stability region is characterized in this paper, was not studied in~\cite{gsk}.  Finally, the results in~\cite{Hong,Han,ritesh,kyuho,awais,yigal,gsk,qiao} are in the context of sub-6GHz wireless networks. These results do not hold in the context of mmWave  networks due to the different propagation characteristics of the mmWave channel as compared to the sub-6GHz channel.

We now briefly review the existing research literature on  user association in mmWave networks; a detailed survey is provided in~\cite{outlook}. The user association problem has been studied with different objectives, such as maximizing energy and spectrum efficiency~\cite{energy},~\cite{power}, spectrum sharing~\cite{sharing}, throughput maximization~\cite{thr}, and maximizing the capacity of the network~\cite{angela}.  In~\cite{fractional}, the user association problem in an mmWave MIMO system was formulated as a mixed integer nonlinear optimization problem, taking into account the load of each BS, and a genetic algorithm was used to solve it. The user association problem in 60-GHz mmWave networks was investigated in~\cite{ocalb} with the objective of balancing the load across different BSs.   A non-convex optimization problem was formulated and a distributed subgradient algorithm based on Lagrangian duality was proposed to solve it. It was shown that the proposed algorithm outperforms the received signal strength based algorithm.  In~\cite{abram}, the user association problem in 60-GHz mmWave networks was modeled as a linear optimization problem with the objective of maximizing the users' throughput. This problem was transformed into the minimum cost flow problem and a solution approach was proposed based on auction algorithms. In the same line of work, a joint user association and relaying problem was studied in~\cite{dacr}. A multi-dimensional integer optimization problem was formulated and an auction based algorithm was proposed as a solution approach.  A mobility aware user association problem was investigated in~\cite{maua}. A Markov chain based association policy was proposed to minimize the number of handovers among BSs.    A cell association problem in hybrid HetNets with tradional macro cells and mmWave small cells was studied in~\cite{duca}. The user association problem in 5G two-tier heterogeneous mmWave networks was investigated jointly with the spectrum allocation problem in~\cite{uawbb}. A cell association problem considering the penalty cost for handovers and reallocation of resources and the intermittent nature of mmWave communication links was investigated in~\cite{ua5g}.  A two-stage BS selection algorithm for heterogeneous mmWave networks was proposed in~\cite{cbss}.

However, there are no proven guarantees on performance for any of the user association schemes proposed in~\cite{ocalb,MDP,energy,cbss,power,sharing,thr,angela,fractional,abram,dacr,maua,duca,uawbb,ua5g}. 
To the best of our knowledge, a characterization of the stability region has not been done in prior work for \emph{any} user association policy for mmWave networks. In contrast, we analytically characterize the stability regions of the SNR based, Throughput based, Load based,  and Mixed association policies for mmWave networks in this paper. Also, LoS/ NLoS propagation and blockage effects in mmWave communication have not been considered in~\cite{ocalb,energy,power,fractional,abram,dacr}. Similarly, blockage and channel fading effects have not been considered in~\cite{angela},~\cite{dacr},~\cite{maua},~\cite{uawbb}. LoS/ NLoS propagation, blockage and channel fading effects can significantly impact link qualities and user-BS association in mmWave networks and are taken into account in the model in this paper. In addition, in the model in~\cite{ocalb,sharing},~\cite{angela,fractional,abram,dacr,maua}, it is assumed that the number of users in the system is constant. In contrast, in the model in this paper, users arrive into and depart from the system over time, as in practice. Also, the user association algorithms proposed in~\cite{ocalb,sharing,thr,angela,fractional,abram,dacr,uawbb,ua5g} are centralized and/ or require the system to solve a complex optimization problem. In contrast, all the user association policies studied in this paper are simple, easy to implement, distributed and online.     

\section{System Model and User Association Policies}
\label{model}
In this section, we describe the considered mmWave network model and channel model in Section~\ref{network model}, and various user association policies in Section~\ref{policy}.
\vspace{-.3cm}
\subsection{Network and Channel Model}
\label{network model}
We consider an mmWave wireless cellular network in which there are  $B$ base stations (BSs); let $\mathcal{B} = \{1, \ldots, B\}$ be the set of BSs. Users arrive into the network according to some random process. The locations of the arriving users may possibly have a non-homogeneous spatial distribution. Each user has a data file to be transmitted. Upon arrival, each user selects a BS to associate with according to some association policy and then starts transmitting its file to the selected BS. 
Each user exits the network after its entire file has been transmitted. We assume that BS $i \in \mathcal{B}$ has $m_i$ radio frequency (RF) chains~\cite{shokri}, where $m_i \geq 1$. Using each RF chain, a BS is able to form one beam; hence, BS $i$ is able to simultaneously exchange data with up to $m_i$ users. However, we assume that each user only has a single RF chain; note that this is often the case in practice since there are constraints on the sizes of user equipment. 

Recall that mmWave communication occurs at high frequencies (30 - 300 GHz), and mmWave channels have propagation characteristics that are completely different from those of channels used in sub-6GHz communication. Since communication in mmWave bands is highly directional and these bands have large bandwidths, mmWave channels are noise-limited~\cite{rangan},~\cite{akdeniz}; hence, we take thermal noise into account,  but neglect interference. Note that similar assumptions have been extensively made in prior research on mmWave communication, e.g., in~\cite{ocalb},~\cite{angela},~\cite{duca},~\cite{shokri,zhu,lin,xu,huang,zhouf}.  Also, mmWave propagation experiences high path loss, low diffraction, high penetration losses and high susceptibility to blocking. Path loss is typically modeled using a non-linear function of the distance between a BS and a user and the path loss exponent~\cite{los}.
Also, due to the blocking effect, a link between a BS and a user may be Line of Sight (LoS) or Non-LoS (NLoS).  

\subsection{User Association Policies}
\label{policy}
Our objective is to  compare the performances of various user association policies in the context of the mmWave network and channel models described in Section~\ref{network model}. In this section, we describe various user association policies; their performance is evaluated via analysis and simulations in the following sections.

Let $\mathcal{X}_i$ be the set of users associated with BS $i$ and $|\mathcal{X}_i| = X_i$. We assume that BS $i$ allocates its available $m_i$ RF chains for equal fractions of time to all the users associated with it. If $X_i \leq m_i$, each user in $\mathcal{X}_i$ is assigned a dedicated RF chain of BS $i$ and hence the throughput achieved by a user in $\mathcal{X}_i$ equals the data rate that it gets with BS $i$. However, if $m_i < X_i$, then each user in $\mathcal{X}_i$ is served by BS $i$ for a fraction $\frac{m_i}{X_i}$ of the time.  Hence, the throughput of a user $j$ associated with BS $i$ at data rate $r_{ij}$ can be expressed as follows:
\begin{equation}
\mathcal{T}_{ij}=\min(m_i,X_i)\frac{r_{ij}}{X_i}. 
\label{thr}
\end{equation}

\textbf{SNR Based Association Policy:} In this policy, an arriving user $j$ associates with the BS $a \in \mathcal{B}$ from which it would get the highest data rate after association (ties are broken at random), \emph{i.e.}:
\begin{equation}
\label{snr}
a = \underset{i \in \mathcal{B}}{\argmax} \; r_{ij}.
\end{equation}
Intuitively, this policy seeks to ensures that communication takes place at high data rates  between BSs and the users that they are associated with.

\textbf{Load Based Association Policy:} In this policy, an arriving user associates with the BS $a \in \mathcal{B}$ which has the minimum number of already associated users, \emph{i.e.,} 
\begin{equation}
a = \underset{i \in \mathcal{B}}{\argmin} \; X_i.
\end{equation}
Intuitively, this policy attempts to balance the user load across the set of BSs, so as to ensure that no BS is overloaded or underutilized.

\textbf{Throughput Based Association Policy:} In this policy, an arriving user $j$ associates with the BS $a \in \mathcal{B}$ from which it would get the highest throughput after association, \emph{i.e.,} 
\begin{equation}
a = \underset{i \in \mathcal{B}}{\argmax} \; \mathcal{T}_{ij},
\end{equation}
where $\mathcal{T}_{ij}$ is given by~\eqref{thr}. 
Intuitively, this policy seeks to ensure that data is transferred at high time-averaged rates (throughput) from users to the BSs that they are associated with. Note that by \eqref{thr}, the throughput of a user is proportional to the  data rate it gets with the BS it is associated with and inversely proportional to the total number of users associated with the BS. Hence, the Throughput based policy seeks to ensure that data is transmitted at high data rates from users to BSs as well as to balance the user load across different BSs.

\textbf{Mixed Association Policy:} For an arriving user $j$, consider the following function:
\begin{equation}
g(i)= \mathcal{T}_{ij}+ \theta r_{ij}, \ i \in \mathcal{B},
\label{mixedeq}
\end{equation}
where $\theta$ is a non-negative constant. The arriving user calculates the value of $g(i), \forall i \in \mathcal{B}$, and the BS, say $a$, for which the value of $g(\cdot)$ is the highest is selected for association, \emph{i.e.}:
\begin{equation}
a = \underset{i \in \mathcal{B}}{\argmax} \; g(i).
\end{equation}
 This is referred to  as the Mixed association policy as it considers both the throughput achieved by the user and the data rate of the user. A motivation for considering this policy is that a similar policy, \emph{RAT}, was found to perform well in the context of user association in sub-6GHz 802.11 WLANs in~\cite{gsk},~\cite{RAWNET}. A typical value of the parameter $\theta$ in \eqref{mixedeq} is $0.2$; this value  was found to result in good performance of the policy in~\cite{gsk},~\cite{RAWNET}.   

Note that all the above policies (SNR Based, Load Based, Throughput Based and Mixed) are simple, easy to implement, distributed and online. To implement these policies, a user only requires knowledge of the qualities of the links from all the BSs in its range  to itself, and the number of users that are already associated with each BS in its range. This information can be easily obtained with the help of reference  signals sent from each BS. 

\section{Stability Analysis of User Association Policies: Two Base Stations}
\label{simplemodel}
In this section, we present a characterization of the stability regions of the user association policies described in Section~\ref{policy}. For tractability, we analyze a simplified version of the model described in Section~\ref{model}; this simplified model is presented in Section~\ref{SSC:simplified:nw:model}.  
A continuous time Markov chain (CTMC) model~\cite{wolfe} is used for the stability analyis; this model and some background are provided in Section~\ref{SSC:CTMC:model}. Finally, the stability analysis of the user association policies is provided in Section~\ref{SSC:stability:analysis}.
\vspace{-.3cm}
\subsection{Simplified Two Base Stations Network Model}
\label{SSC:simplified:nw:model}
\begin{figure}[!hbt]
\centering
\resizebox{.9\columnwidth}{!}{\includegraphics{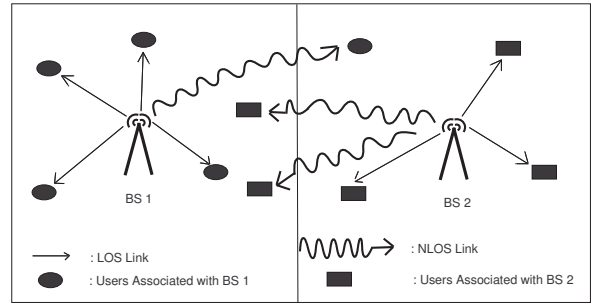}}
\caption{Simplified model with two BSs located at the centres of two square hotspots. The link between a user and the BS in the same square (respectively, other square) is LoS (respectively, NLoS) as shown in the figure by a solid (respectively, wavy) arrow.}
\label{simple:p1}
\end{figure}

We consider an mmWave network with two BSs $\{1,2\}$ located at the centres of two square hotspots as shown in Fig.~\ref{simple:p1}. For example, the squares may be two conference halls separated by a partition. Users arrive into the network according to a Poisson process with parameter $\lambda$. Each user has a data file to be transmitted. Suppose the sizes of the files of different users are independent and  exponentially distributed random variables with parameter $\mu$. Upon arrival, each user associates with one of the two BSs, selected according to some association policy. After association, the user starts transmitting its file to the BS it associated with, and departs from the system once its file has been transmitted completely. 

For tractability, we assume that the propagation link between a user and a BS can be in one of two possible states: LoS and NLoS; similar models have been used in~\cite{cbss},~\cite{angela},~\cite{los},~\cite{mai}. In particular, the link between a user and the BS at the centre of the square in which the user is located is LoS and the link between the user and the other BS is NLoS.  Let the data rates supported by BSs over NLoS and LoS links be $c_1$ and $c_2$ respectively, where $c_2 > c_1$. Then, an arriving user can associate with a BS at a rate either $c_1$ or $c_2$. A rate vector $\mathcal{R}=(r_{1j},r_{2j})$ represents that a user $j$ can associate with BS 1 (respectively, BS 2) at rate $r_{1j}$ (respectively,  $r_{2j}$).  Note that $(c_1,c_2)$ and $(c_2,c_1)$ are the two possible rate vectors for an arriving user. The rate vector is $(c_2,c_1)$ (respectively, $(c_1,c_2)$) if the user arrives in the left (respectively, right) square. For tractability, we assume that the rate vector of a given user remains the same from its instant of arrival into the network until its departure. Suppose each arriving user has a rate vector $(c_2,c_1)$ with probability $p_1$ and $(c_1,c_2)$ with probability $p_2 = 1 - p_1$. Then the pair $(\lambda, p_1)$ parametrizes the arrival process in the system.     

\vspace{-.3cm}
\subsection{CTMC Model Used for Stability Analysis, and Background} 
\label{SSC:CTMC:model}
Under the model described in Section~\ref{SSC:simplified:nw:model}, users arrive into the system, transmit their files and depart from the system. Hence, the system can be modeled as a \emph{queueing system}~\cite{wolfe}. Recall that continuous time Markov chains (CTMC)~\cite{wolfe} have been extensively used in prior research to model queueing systems, including those that arise in the field of communication networks~\cite{yue,petrov,tsang,leung,xing,jiang}. A CTMC model can also be used in our context to effectively analyze the system stability.

In particular, under the model described in Section~\ref{SSC:simplified:nw:model}, for an arrival process parametrized by a given pair $(\lambda, p_1)$, a fixed user association policy induces a CTMC~\cite{wolfe}. A state, say $s$, of this CTMC is given by:
\begin{equation}
s = \left( \begin{array}{cc}
X_{11} & X_{12} \\
X_{21} & X_{22}
\end{array} \right),
\end{equation}   
where for $i \in \{1,2\}$ and $r \in \{1,2\}$, $X_{ir}$ denotes the number of users associated with BS $i$ at rate $c_r$. We use the concepts of \emph{recurrent} and \emph{transient} CTMCs~\cite{wolfe} to model the stability and instability of the system. Under a given user association policy, the system is stable (respectively, unstable) if the induced CTMC is positive recurrent (respectively, null recurrent or transient)~\cite{wolfe}. Intuitively, if the induced CTMC is positive recurrent (respectively, null recurrent or transient), the number of users in the system will tend to be small most of the time (respectively, tend to increase without bound as time progresses)~\cite{wolfe}, and hence the network will provide good (respectively, poor) Quality of Service (QoS). The set of all pairs $(\lambda, p_1)$ for which the system is stable is referred to as the \emph{stability region} of the user association policy. Also, the set of all pairs $(\lambda, p_1)$ for which there exists a user association policy that stabilizes the system is referred to as the \emph{capacity region} of the system.  Note that the stability region of every user association policy is a subset of the capacity region.

Next, the transition rate~\cite{wolfe} out of a state $s$ is the reciprocal of the expected amount of  time for which the system stays in the state $s$.  If $X_{i1} + X_{i2} > 0$, then by~\eqref{thr}, the throughput of each user associated with BS $i$ at rate $c_1$ (respectively, $c_2$) is $\min(m_i,(X_{i1}+X_{i2}))\frac{c_1}{X_{i1}+X_{i2}}$ (respectively, $\min(m_i,(X_{i1}+X_{i2}))\frac{c_2}{X_{i1}+X_{i2}}$). So if ($X_{11} \ne 0$ or $X_{12} \ne 0$) and ($X_{21} \ne 0$ or $X_{22} \ne 0$), then
the total transition rate out of state $s$ is given by:

\begin{eqnarray}
\nu(s) =  
 &\lambda + \mu \min(m_1,(X_{11}+X_{12}))\frac{X_{11}c_1+X_{12}c_2}{X_{11}+X_{12}}  \nonumber \\
& + \mu \min(m_2,(X_{21}+X_{22}))\frac{X_{21}c_1+X_{22}c_2}{X_{21}+X_{22}}. 
\label{EQ:aj:argmax}
\end{eqnarray}
The first term in the above expression is the arrival rate of users and the second (respectively, third) term is the departure rate from BS $1$ (respectively, BS $2$). If there are no associated users with BS $1$ (respectively, BS $2$), \emph{i.e.,} $X_{11}=X_{12}=0$ (respectively, $X_{21}=X_{22}=0$), then the transition rate is given by \eqref{EQ:aj:argmax} with the second (respectively, third) term removed from the RHS. 

Next, we consider the embedded discrete time Markov chain (EDTMC)~\cite{wolfe} corresponding to the above CTMC. An EDTMC is a DTMC obtained by sampling a CTMC at the transition instants~\cite{wolfe}. The transition probabilities of this  EDTMC are as follows.  If $X_{ir} \ne 0$, then the probability of transition from state $s$ to the state reached when the departure of a user associated with BS $i$ at rate $c_r$ occurs is given by:
\begin{equation}
\min(m_i, X_{i1}+X_{i2})\frac{\frac{\mu X_{ir} c_r}{X_{i1}+X_{i2}}}{\nu(s)}.
\end{equation} 
Recall that the rate vector of a new user is $(c_2,c_1)$ (respectively, $(c_1,c_2)$) with probability $p_1$ (respectively, $p_2$). Let $p_{total}(s,i,r)$ be the sum of the  probabilities of all the rate vectors of a new user for which the user joins BS $i$ at rate $c_r$.  Then, the transition probability from state $s$ to the state reached when a user joins BS $i$ at rate $c_r$  is given by:
\begin{equation}
\frac{\lambda p_{\mathit{total}}(s,i,r)}{\nu(s)}.
\end{equation}
For example, under the SNR based association policy, the transition probability from state $s$ to the state reached when a new user associates with BS $1$ (respectively, BS $2$) at rate $c_2$ is $\frac{\lambda p_1}{\nu(s)}$ (respectively, $\frac{\lambda p_2}{\nu(s)}$).

Now, consider a discrete time Markov chain (DTMC) with state space $S$. Let the transition probability  from state $m$ to state $n$ be $p_{mn}$.     Given a function $V: S \rightarrow \Re$, the \emph{drift}~\cite{asmussen} in state $m$ is defined as follows:
\begin{equation}
\Delta V(m) = \sum_{n \in S} p_{mn} (V(n)-V(m)).  
\end{equation}
We now state two theorems from~\cite{asmussen}, which we use to prove stability and instability of the Markov chains induced under various user association policies.

\begin{theorem}
\label{stable}
Consider an irreducible DTMC and let $S_0$ be a finite set such that $S_0 \subset S$, where $S$ is the state space. Suppose for some function $V: S \rightarrow \Re$ and for some 
$\epsilon > 0$:
\begin{eqnarray}
\inf_{x \in S} V(x) & > & - \infty, \label{stable1} \\ 
\sum_{n \in S} p_{mn} V(n) & < & \infty, \ \forall m \in S_0, \label{stable2} \\
\Delta V(m) & \le &  - \epsilon, \ \forall m \notin S_0. \label{stable3}
\end{eqnarray}
Then the DTMC is positive recurrent. 
\end{theorem}

\begin{theorem}
\label{unstable}
Consider an irreducible DTMC and let $S_0$ be a finite set such that $S_0 \subset S$, where $S$ is the state space. Let $V: S \rightarrow \Re$ be a function such that:
\begin{eqnarray}
\Delta V(m) & \ge & 0, \ \forall m \notin S_{0}, \label{unstable1} \\
 V(n) & > & V(m) \mbox{ for some } n \notin S_0 \nonumber \\
               & &   \mbox{ and all } m \in S_0, \label{unstable2} \\
 \inf_{x \in S} V(x) &  > & - \infty, \label{unstable3} \\
 \sum_{n \in S} p_{mn} \left| V(n) - V(m) \right| & \le & B, \ \forall m \in S, \label{unstable4}
\end{eqnarray}
for some $B < \infty$. Then the DTMC is null recurrent or transient. 
\end{theorem}

To prove the stability or instability of a Markov chain using Theorem~\ref{stable} or~\ref{unstable}, the key challenge is to find a function $V(\cdot)$ that satisfies the conditions in the theorem. Such a function $V(\cdot)$ is known as a \emph{Lyapunov function}~\cite{asmussen}. Also, using \eqref{EQ:aj:argmax}, it is easy to show that the total transition rate out of state $s$, \emph{i.e,} $\nu(s)$, is lower as well as upper bounded by constants. Therefore, the original CTMC is positive recurrent (respectively, null recurrent or transient)  iff the corresponding EDTMC is positive recurrent (respectively, null recurrent or transient)~\cite{eylem}. So henceforth, we focus on proving the stability or instability of the EDTMC using Theorem~\ref{stable} or~\ref{unstable}; the stability or instability of the original CTMC follows.  

\subsection{Stability Analysis}
\label{SSC:stability:analysis}
We now present a stability analysis of different user association policies. Throughout this section, for algebraic simplicity, $c_1$ is normalized to 1.
The proofs of all the analytical results in this section are relegated to the Appendix. 

The following result characterizes the capacity region of the system for the special case $m_1=m_2=1$. 
\begin{theorem}
\label{thm:capacity}
For $m_1=m_2=1$, the system can be stabilized iff $\lambda < \lambda^*$, where\\ 
\begin{math}
 \lambda^*  =\frac{\mu c_2 (1+c_2)}{\max(p_1,p_2)c_2 + \min(p_1,p_2)}.
\end{math}
\end{theorem}

The following result characterizes the stability region of the SNR based policy. 
\begin{theorem}
\label{thm:SNR}
Under the SNR based policy, the system is stable iff
\begin{math}
\lambda < \mu c_2  \min \left(\frac{m_1}{p_1},\frac{m_2}{p_2} \right).
\end{math}
\end{theorem}

From Theorem~\ref{thm:SNR}, it can be seen that as the number of RF chains per BS increases, the stability region of the SNR policy expands, which is consistent with intuition, since the capacity of a BS to serve users is higher when it has more RF chains.  
Also, from Theorems~\ref{thm:capacity} and~\ref{thm:SNR}, it follows that if $m_1=m_2=1$ and $p_1=p_2= 0.5$, then the SNR based policy is maximally stable. However, for $p_1 \neq 0.5$, the SNR based policy is sub-optimal. Intuitively, this is because when $m_1 = m_2 = 1$, the capacities of the two BSs to serve users are equal; also, when $p_1 = p_2 = 0.5$, the SNR based policy splits the arriving user load equally across the two BSs and all users are served at the higher rate $c_2$, which results in the SNR based policy being maximally stable. However, when $p_1 > 0.5$ (respectively, $p_1 < 0.5$), BS 1 (respectively, BS 2) is overloaded and BS 2 (respectively, BS 1) is underutilized under the SNR based policy, which results in the policy being sub-optimal.      

The following result characterizes the stability region of the Load based policy. 
\begin{theorem}
\label{thm:load}
(a) If $c_2=2$, then under the Load based policy, the system is unstable if $\lambda \geq \frac{2\mu(m_1+m_2)}{1+\min(p_1,p_2)}$. (b) If $c_2=2$ and  $m_1=m_2=1$, then under the Load based policy, the system is stable if $\lambda < \mu$.
\end{theorem}

Note that as the number of RF chains per BS increases, the bound, $\frac{2\mu(m_1+m_2)}{1+\min(p_1,p_2)}$, on the stability region of the Load based policy provided by Theorem~\ref{thm:load}(a) increases, which is consistent with intuition, since the capacity of a BS to serve users is higher when it has more RF chains. 

It is difficult to characterize the stability regions of the Throughput based and Mixed policies for arbitrary values of $m_1,m_2$ and $p_1$. However, we have obtained stability results for the two policies for the special cases of  completely non-uniform $(p_1=1)$ and completely uniform $(p_1=0.5)$ spatial distributions of arriving users. 
\begin{theorem}
\label{thm:selfish}
Suppose $c_2 = 2$. (a) If $p_1=1$ and $m_1 = m_2 = 1$, then under the Throughput based policy, the system is stable for all $\lambda < 3\mu$.
(b) If $p_1=m_1=1$ and $m_2=2$, then under the Throughput based policy, the system is stable for all $\lambda < 4 \mu$.
(c) If $p_1=0.5, m_1 = m_2 = 1$, then under the Throughput based policy, the system is stable  for all $\lambda < \frac{4}{3} \mu$. 
\end{theorem}

Note that in Theorem~\ref{thm:selfish}, in the case in which $p_1 = 1$, $c_2 = 2$ and $m_1 = 1$, when $m_2$ increases from $1$ to $2$, we are able to guarantee a larger stability region for the Throughput based policy. This is consistent with intuition-- when the number of RF chains at BS 2 increases, it can serve more users simultaneously, which increases its stability region.

\begin{theorem}
\label{thm:heuristic}
Suppose $\theta =0.2$. (a) Let $m_1=m_2=1$. For $p_1=0.5$ and arbitrary $c_2$, under the Mixed policy, the system is stable for $\lambda < 2 \mu c_2$. If $p_1=1$ and $c_2=2$, then under the Mixed policy, the system is stable for $\lambda   <   (\sqrt{3} +1) \mu$.
(b) If $p_1=m_2=1$ and $m_1=c_2=2$, then under the Mixed policy, the system is stable for $\lambda < (3+\sqrt{3})\mu.$
\end{theorem}

In Theorem~\ref{thm:heuristic}, the parameter $\theta$ in \eqref{mixedeq} is assumed to be $0.2$ for concreteness. Results similar to Theorem~\ref{thm:heuristic} can be shown for arbitrary values of $\theta > 0$.   
Next, note that in Theorem~\ref{thm:heuristic}, in the case in which $p_1 = 1$, $c_2 = 2$ and $m_2 = 1$, when $m_1$ increases from $1$ to $2$, we are able to guarantee a larger stability region for the Mixed policy. This is consistent with intuition-- when the number of RF chains at BS 1 increases, it can serve more users simultaneously, which increases its stability region.  Next, from Theorems~\ref{thm:capacity},~\ref{thm:SNR}  and~\ref{thm:heuristic}, it can be seen that for a uniform spatial distribution of arriving users (\emph{i.e.}, $p_1 = 0.5$) and $m_1=m_2=1$, the Mixed and SNR based policies are both maximally stable; if, additionally, $c_2 = 2$, then by Theorem~\ref{thm:load}(a), the Load based policy is sub-optimal. 
\begin{remark}
As the above analysis and the proofs in the Appendix show, even though in our simplified model, it is assumed for tractability that there are two BSs and two possible data rates, the analysis of the simplified model is non-trivial. Also, our simplified model takes into account several key aspects of the user association problem and our analysis provides a number of insights as explained above. Nonetheless, in Section~\ref{gen_case}, we have generalized several of  the results obtained via the above stability analysis to a model in which there are an arbitrary number of BSs and two possible data rates. An approximate analysis of the case in which there are an arbitrary number of BSs and arbitrary number of possible data rates is a direction for future research.
\end{remark}

\section{Stability Analysis of User Association Policies: Arbitrary Number of Base Stations}
\label{gen_case}
In this section, we generalize the results obtained via stability analysis  of the simplified two BS model in Section~\ref{SSC:stability:analysis} to the case where there are an arbitrary number of BSs. The system model and CTMC model used for the arbitrary number of BS case are provided in Section~\ref{gen_model} and the stability region analysis of different user association policies for the arbitrary number of BS case is provided in Section~\ref{gen_stability}. 
\subsection{Network Model and CTMC Model}
\label{gen_model}
Consider the network model described in Section~\ref{SSC:simplified:nw:model} with the following changes. There are $B \geq 2$ BSs in the mmWave network. As before, we assume that the propagation link between a user and a BS can be in one of two possible states: LoS and NLoS. Let the data rates supported by BSs over NLoS and LoS links be $c_1$ and $c_2$, respectively, where $c_2 > c_1$. If a user $j$ located in the cell of BS $i$ associates with BS $i$ (respectively, a BS other than $i$), then the link is LoS (respectively, NLoS) and the supported rate is $c_2$ (respectively, $c_1$). A rate vector $\mathcal{R} = (r_{1j},\ldots, r_{ij},\ldots, r_{Bj} )$ represents that user $j$ can associate with
BS $i$ with rate $r_{ij}, i \in \{1,\ldots,B\}$. Note
that there are $B$ possible rate vectors
for an arriving user. Let $\mathcal{R}_i$ be the rate vector of a user that arrives in the cell of BS $i \in \{1,\ldots,B\}$; then the $i$'th element of the rate vector $\mathcal{R}_i$ is $c_2$ and all the other elements are $c_1$. Suppose each arriving user
has rate vector $\mathcal{R}_i$
with probability $p_i$, for $i \in \{1, \ldots, B\}$, and let $\textbf{P}=\{p_1,\ldots,p_B\}$ be such that $\sum_{i=1}^{B}p_i =1$. Then the pair ($\lambda$, $\textbf{P}$) parametrizes
the arrival process in the system.

Similar to the CTMC model in Section~\ref{SSC:CTMC:model}, a state, say $s$, of
the induced CTMC is given by:
\begin{equation}
s=\begin{pmatrix}
X_{11} & X_{12}\\
\vdots & \vdots\\
X_{B1} & X_{B2}
\end{pmatrix}
\end{equation}
where for $i\in\{1,\ldots,B\}$ and $r\in \{1,2\}$, $X_{ir}$ denotes the
number of users associated with BS $i$ at rate $c_r$. Also, if $X_{i1}+X_{i2}>0$, then by (\ref{thr}), the throughput of each user associated with BS $i$ at rate $c_1$ 
(respectively, $c_2$) is $\min (m_i, (X_{i1}+X_{i2}))\frac{c_1}{X_{i1}+X_{i2}}$ (respectively, $\min (m_i, (X_{i1}+X_{i2}))\frac{c_2}{X_{i1}+X_{i2}}$). So if $X_{i1}+X_{i2}>0, \forall i\in \{1,\ldots,B\}$, then the total transition rate out
of state $s$ is given by:
\begin{equation}
\nu(s)=\lambda+  \sum_{i=1}^{B} \mu \min (m_i, (X_{i1}+X_{i2}))\frac{X_{i1}c_1+X_{i2}c_2}{X_{i1}+X_{i2}}. \label{Eq8}
\end{equation}
The first term in the above expression is the arrival rate of
users and the $i$'th term of the summation is the departure
rate from BS $i$. If there are no associated
users with BS $i$, \emph{i.e.}, $X_{i1}+X_{i2}=0$, for some values of $i \in \{1, \ldots, B\}$, then the transition rate is given
by (\ref{Eq8}) with the corresponding terms of the summation removed from the RHS.     

\subsection{Stability Analysis}
\label{gen_stability}
We now present a stability analysis of different user association policies for the network model described in Section~\ref{gen_model}. Throughout this section, for algebraic simplicity,
$c_1$ is normalized to 1. The proofs of all the analytical results
in this section are relegated to the Appendix.

The following result characterizes the capacity region of the system in a special case.
\begin{theorem}
	Let $m_i=1, \forall i\in\mathcal{B}$. 
	 Suppose $p_i=\frac{1}{K}\, \forall i \in \{1,\ldots,K\}$ and $p_i=0 \, \forall i \in \{K+1, \ldots ,B\}$ for some $K \in \{1,\ldots, B\}$. The system can be stabilized iff $\lambda < \lambda_o$, where $\lambda_o=(B-K+K c_2)\mu$.
\label{cap_gen}
\end{theorem}

From Theorem~\ref{cap_gen}, it can seen that if $m_i=1\, \forall i \in \mathcal{B}$ and hence the capacities of the BSs to serve users are equal,  and a user arrives in cell $i$, where $i$ is selected uniformly at random from the range $\{1, \ldots, K\}$ independently for each user, then the \emph{capacity region expands as the value of $K$ increases}. Intuitively, this is because the user load can be uniformly distributed across more BSs, while ensuring that each user communicates at the higher data rate $c_2$ with the BS it is associated with.  

The following result characterizes the stability region of the
SNR based policy.
\begin{theorem}
\label{snr2}
Under the SNR based policy, the system is stable iff $\lambda <\mu c_2 \min_{i\in \mathcal{B}}(\frac{m_i}{p_i})$.
\end{theorem}

Trends and intuition similar to those explained for Theorems~\ref{thm:capacity} and~\ref{thm:SNR} (see the paragraph after Theorem~\ref{thm:SNR}) hold for Theorems~\ref{cap_gen} and~\ref{snr2}.

The following result characterizes the stability region of the Load based policy.
\begin{theorem}
\label{load2}
(a) If $c_2=2$, then under the Load based policy, the system is unstable if $\lambda \ge \frac{2 \mu \sum_{i=1}^{B}m_i}{2-\max(p_1, p_2, \ldots, p_B)}$. (b) If $c_2=2$ and $m_i=1$, $\forall  i \in \mathcal{B}$, then under the Load based policy, the system is stable if $\lambda < \mu$.
\end{theorem}

Note that as the number of RF chains per BS increases,
the bound, $\frac{2 \mu \sum_{i=1}^{B}m_i}{2-\max(p_1, p_2, \ldots, p_B)}$, on the stability region of the Load
based policy provided by Theorem~\ref{load2}(a) increases; the intuition for this trend is similar to that provided in the paragraph after Theorem~\ref{thm:load}.

\begin{remark}
Although in Sections~\ref{SSC:stability:analysis} and~\ref{gen_stability}, we have mathematically analyzed the performances of the four user association policies in terms of their \emph{stability regions}, in Section~\ref{simulation}, via simulations, we show that those policies which perform well in terms of the stability region also perform well in terms of other performance metrics widely used in the networking literature such as \emph{average throughput, average delay and fairness} and vice versa.
\end{remark}

\section{Simulations}
\label{simulation}
We evaluate the performances of the user association policies described in Section~\ref{policy} via simulations in this section. In Section~\ref{SSC:simplified:model:simulations}, we investigate as to how tight the bounds on stability regions obtained via analysis in Sections~\ref{SSC:stability:analysis} and~\ref{gen_stability} are. In Section~\ref{SSC:simulations:realistic}, we compare the performances of different user association policies via simulations using a realistic network model.

\subsection{Simulation Results for the Models Described in Sections~\ref{SSC:simplified:nw:model} and~\ref{gen_model}}
\label{SSC:simplified:model:simulations}

\begin{figure}[!hbt]
\begin{minipage}[b]{0.484\columnwidth}
\begin{subfigure}[t]{1in}
\centering
\includegraphics[width=4.7cm, height=3cm]{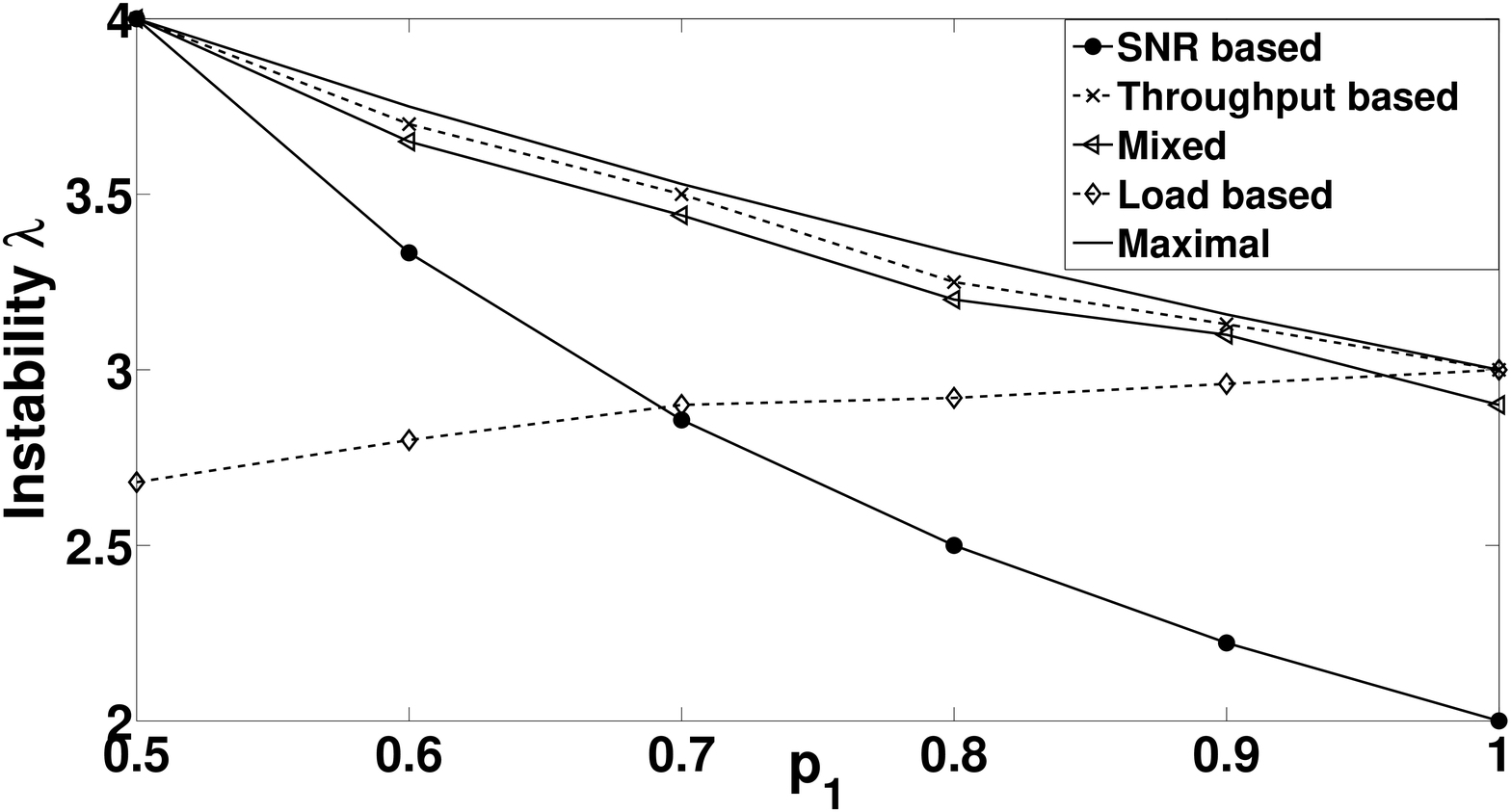}
\caption{\hspace*{-4.5cm}}\label{2a}
\end{subfigure}
\end{minipage}
\hspace{0.2mm}
\begin{minipage}[b]{0.484\columnwidth}
\begin{subfigure}[t]{1in}
\centering
\includegraphics[width=4.7cm, height=3cm]{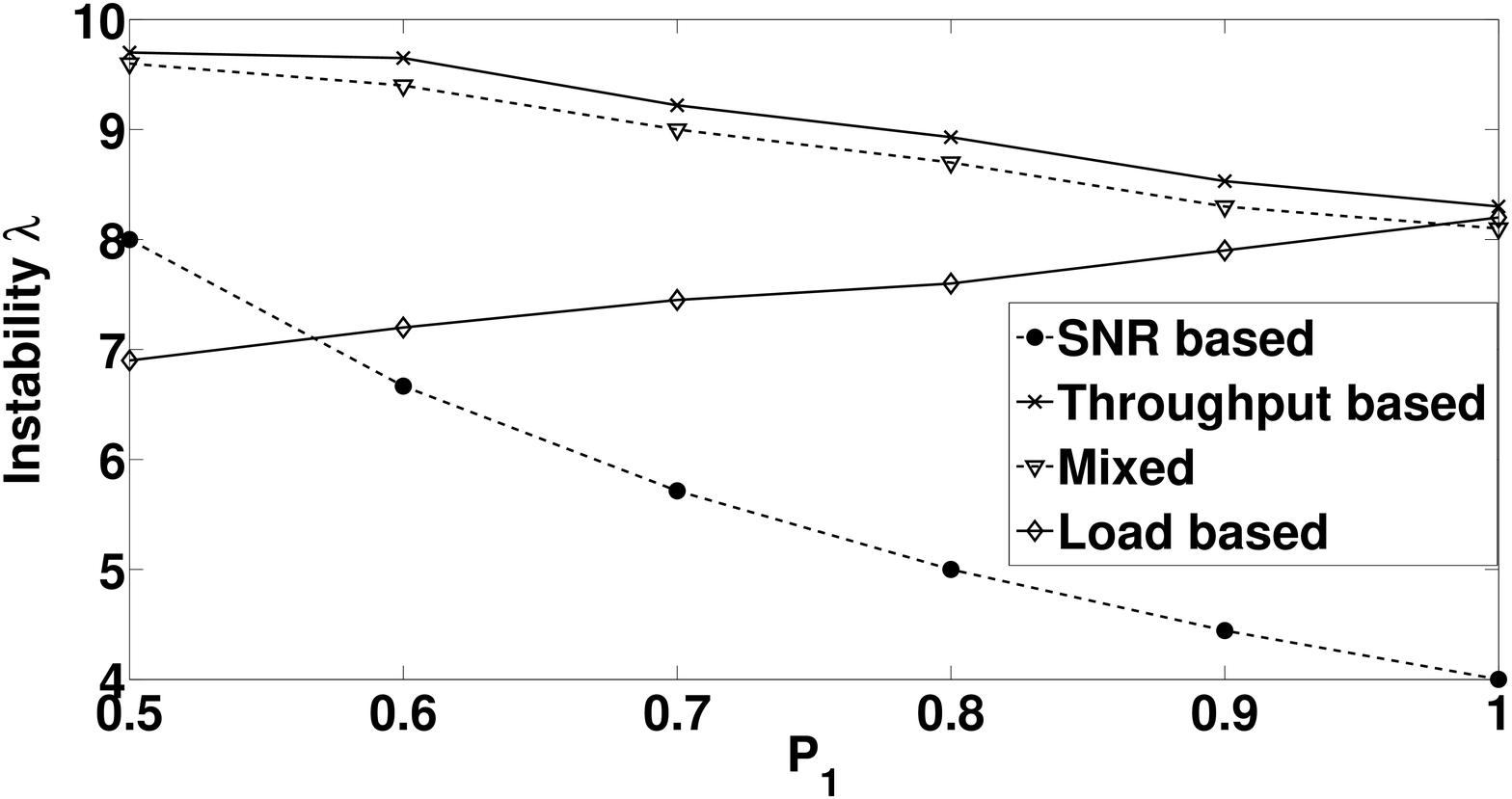}
\caption{\hspace*{-4.5cm}}\label{2b}
\end{subfigure}
\end{minipage}
\centering
\caption{The plot on the left (respectively, right) shows the minimum value of $\lambda$ for which instability is  observed under different user association policies vs probability $p_1$ for the model with two BSs described in Section~\ref{SSC:simplified:nw:model}. The parameter values $m_1 = m_2 = 1$ (respectively, $m_1=2, m_2=3$) and $\mu = 1$ are used. The plot on the left also shows $\lambda^*$ in Theorem~\ref{thm:capacity}, which is the maximal instability threshold for any user association policy.   }
\label{2}
\end{figure}

\begin{figure}[!hbt]
\begin{minipage}[b]{0.484\columnwidth}
\begin{subfigure}[t]{1in}
\centering
\includegraphics[width=4.7cm, height=3cm]{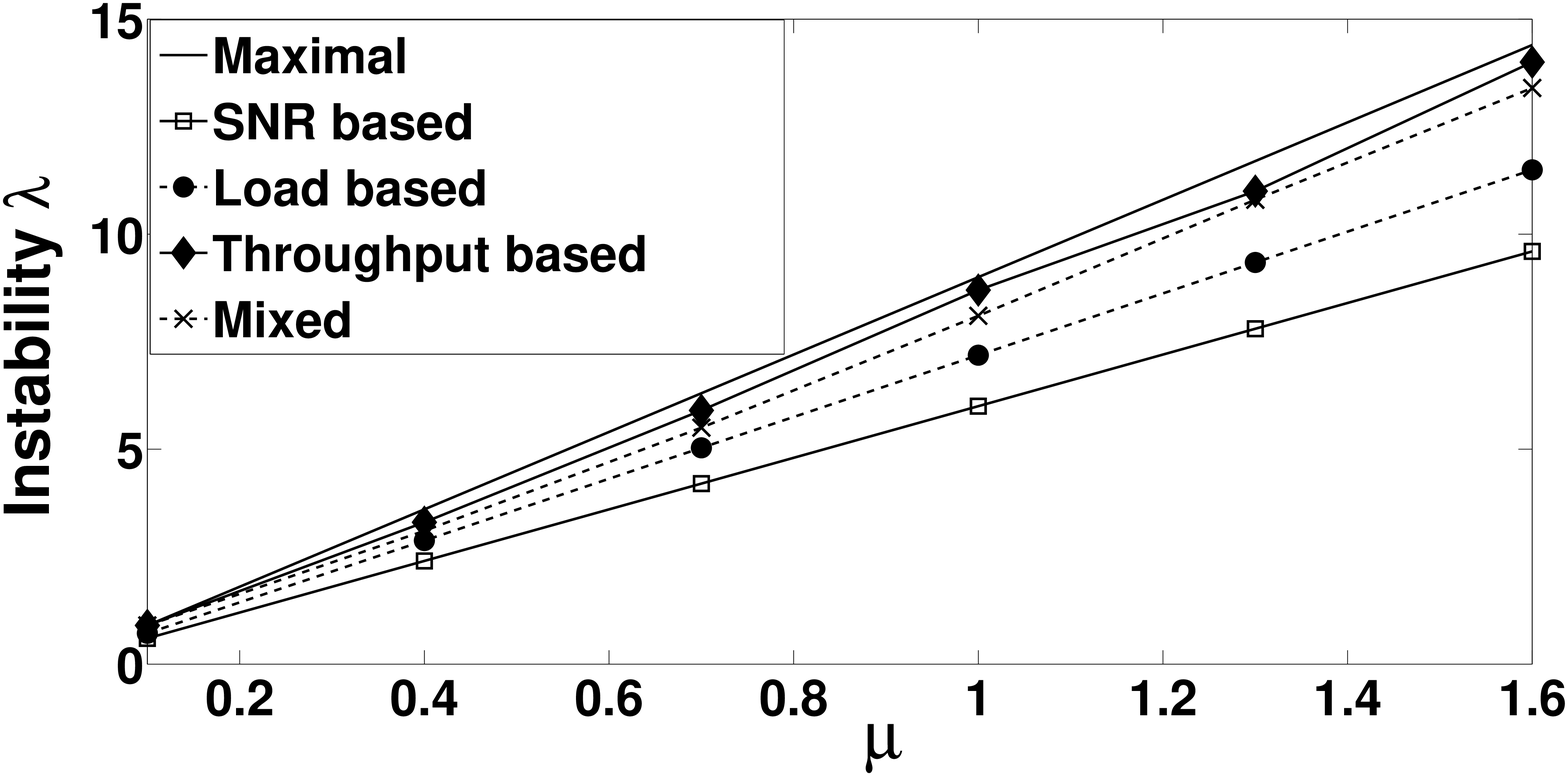}
\caption{\hspace*{-4.5cm}}\label{ga}
\end{subfigure}
\end{minipage}
\hspace{0.2mm}
\begin{minipage}[b]{0.484\columnwidth}
\begin{subfigure}[t]{1in}
\centering
\includegraphics[width=4.7cm, height=3cm]{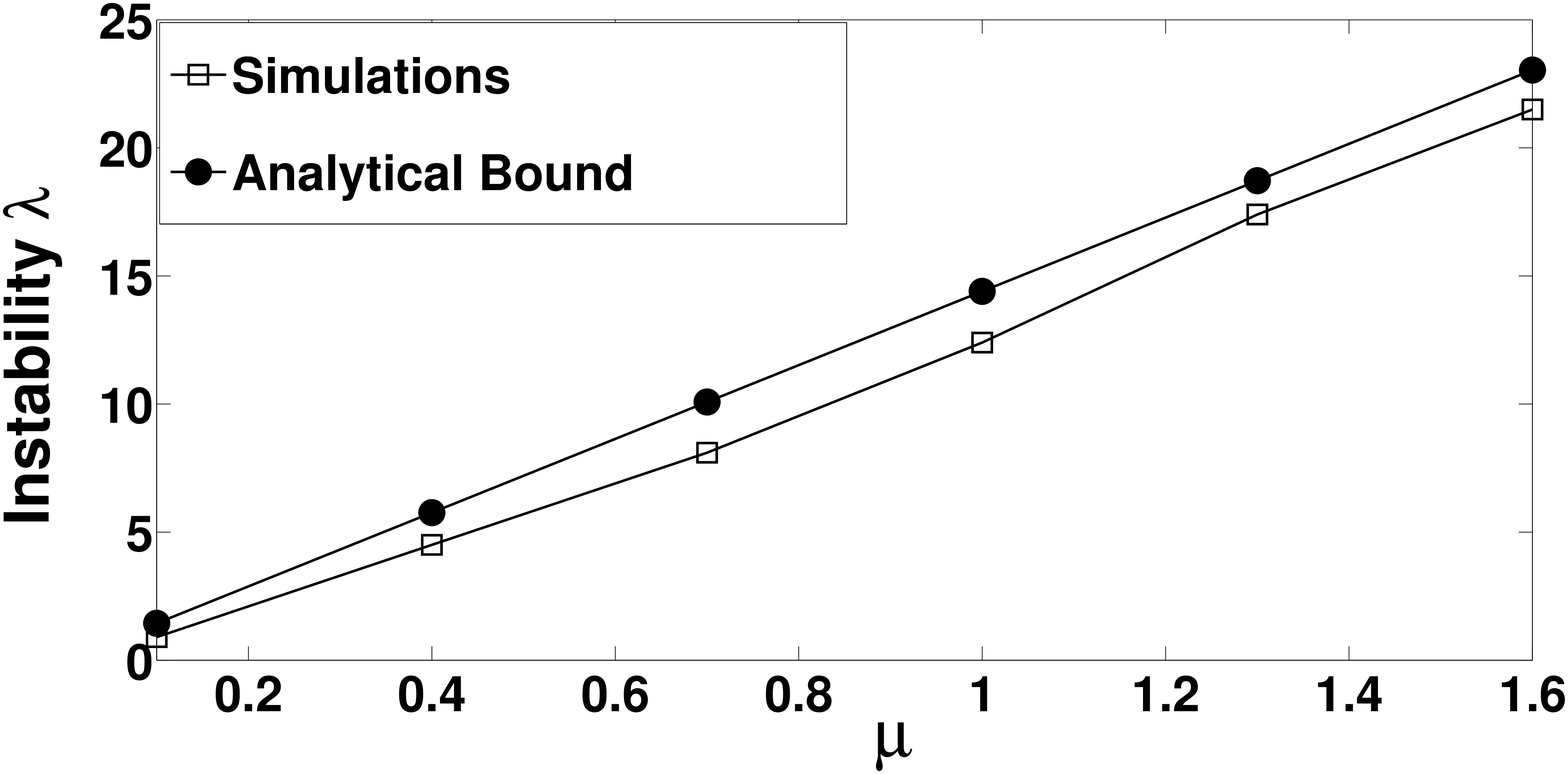}
\caption{\hspace*{-4.5cm}}\label{gb}
\end{subfigure}
\end{minipage}
\centering
\caption{Both the plots are for the model with an arbitrary number of BSs described in Section~\ref{gen_model} with parameter values $B=6$ and $K=3$. The plot on the left shows the minimum value of $\lambda$ for which instability is  observed under different user association policies and the maximal instability threshold for any user association policy, \emph{i.e.,} $\lambda_o$ in Theorem~\ref{cap_gen} vs $\mu$ considering that $m_i=1 \,\forall i \in \mathcal{B}$. The plot on the right shows the minimum value of $\lambda$ for which instability is observed under the Load based policy and the analytical bound shown on it in Theorem~\ref{load2}(a) vs $\mu$ considering that each of the odd numbered (respectively, even numbered) BSs has two (respectively, three) RF chains.}
\label{gen}
\end{figure}

Consider the network models described in Sections~\ref{SSC:simplified:nw:model} and~\ref{gen_model} with parameter values $c_1=1$ and $c_2=2$. 
First, for the two BSs model described in Section~\ref{SSC:simplified:nw:model}, the instability threshold, \emph{i.e.}, the minimum value of $\lambda$ at which the system becomes unstable, under different user association policies has been plotted versus $p_1$ in the two plots of  Fig.~\ref{2} for different values of $m_1$ and $m_2$. Next, consider the model with an arbitrary number of BSs described in Section~\ref{gen_model}, and suppose  a new user arrives uniformly at random in the cell of any of the first  $K$ BSs out of the $B$ BSs, where $K \in \{1,\ldots,B\}$.  For this model, the instability thresholds under different user association policies have been plotted versus $\mu$ in Fig.~\ref{gen}(\subref{ga}) for $B=6,\,K=3$ and $m_i=1 \,\forall i \in \mathcal{B}$.
In Fig.~\ref{2} (respectively,~\ref{gen}(\subref{ga})), the Throughput based policy outperforms the SNR based, Load based and Mixed policies for all values of $p_1$ (respectively, $\mu$); also, Fig.~\ref{2}(\subref{2a}) (respectively,~\ref{gen}(\subref{ga})) shows that for $m_i=1 \,\forall i \in \mathcal{B}$, the Throughput based policy is nearly maximally stable for all values of $p_1$ (respectively, $\mu$) considered. 

\begin{figure}[!hbt]
\begin{minipage}[b]{0.484\columnwidth}
\begin{subfigure}[t]{1in}
\centering
\includegraphics[width=4.7cm, height=3cm]{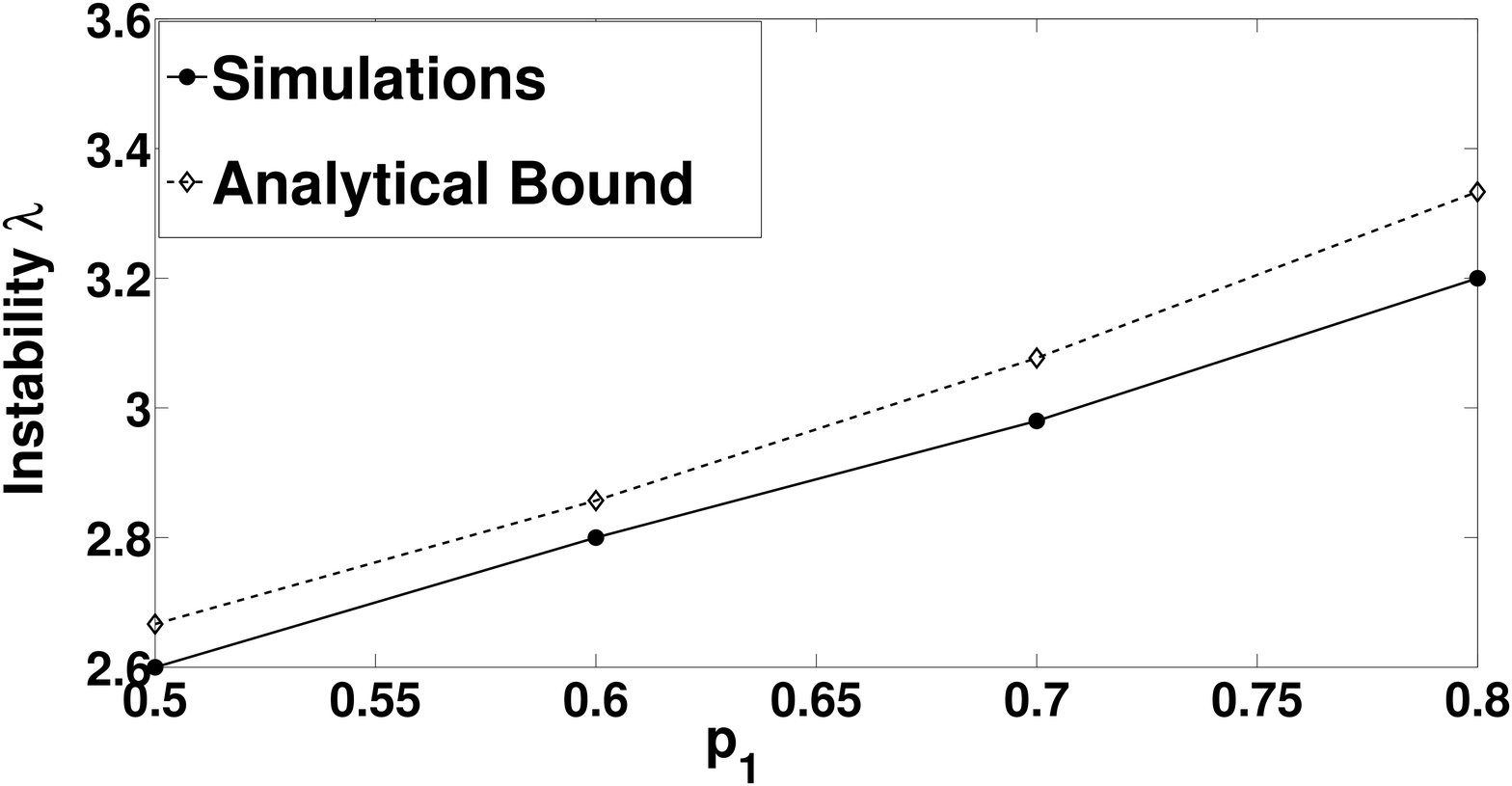}
\caption{\hspace*{-4.5cm}}\label{3a}
\end{subfigure}
\end{minipage}
\hspace{0.2mm}
\begin{minipage}[b]{0.484\columnwidth}
\begin{subfigure}[t]{1in}
\centering
\includegraphics[width=4.7cm, height=3cm]{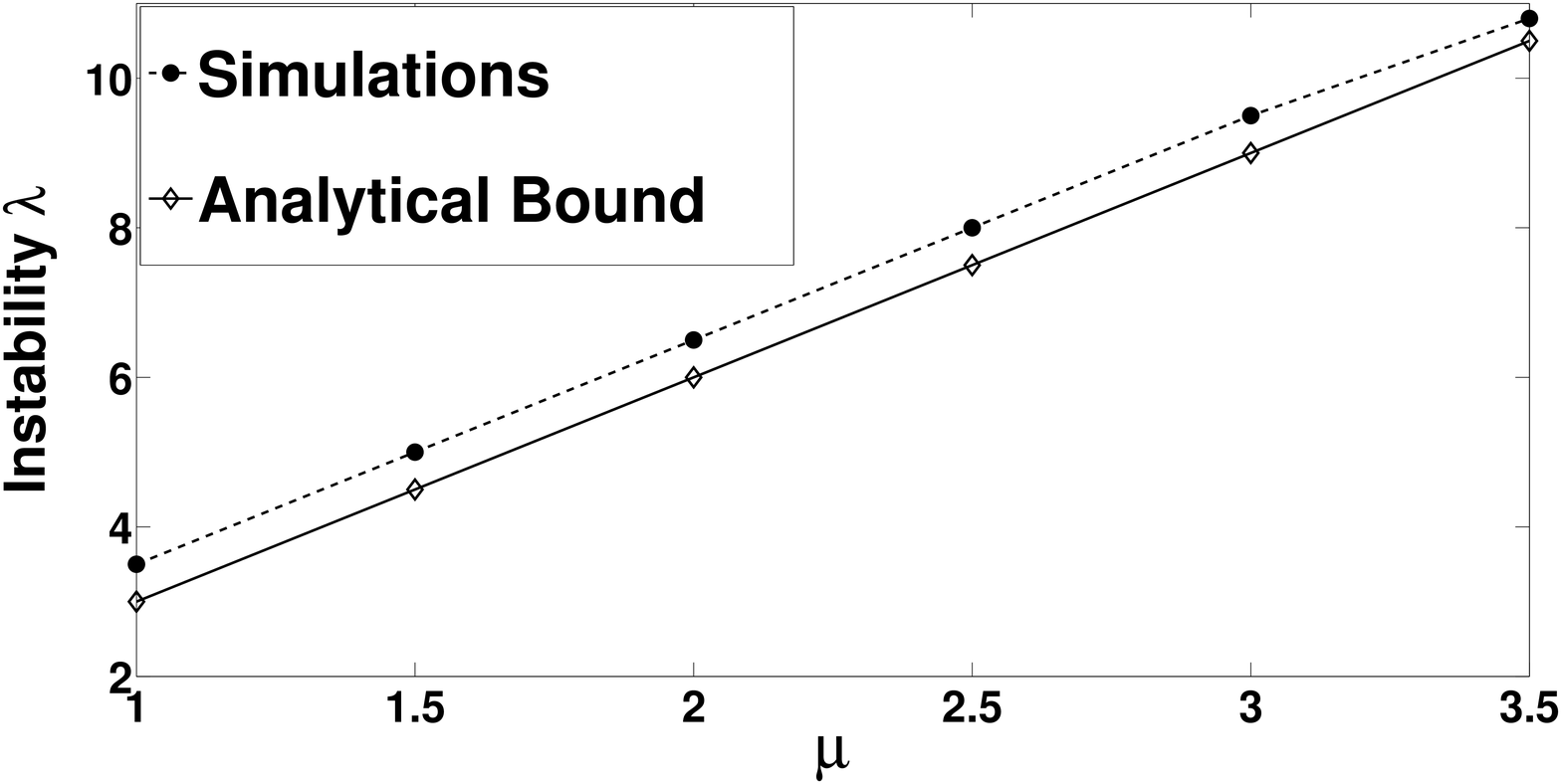}
\caption{\hspace*{-4.5cm}}\label{3b}
\end{subfigure}
\end{minipage}
\centering
\caption{The plot on the left (respectively, right) shows the minimum value of $\lambda$ for which instability is observed under the Load based policy  (respectively, Throughput based policy) and the analytical bound shown on it in Theorem~\ref{thm:load}(a) (respectively, Theorem~\ref{thm:selfish}(a))
vs probability $p_1$ (respectively, parameter $\mu$). The parameter values  $\mu = 1$ (respectively, $p_1=1$) and $m_1 = m_2 = 1$ are used.}
\label{3}
\end{figure}

\begin{figure}[!hbt]
\begin{minipage}[b]{0.484\columnwidth}
\begin{subfigure}[t]{1in}
\centering
\includegraphics[width=4.7cm, height=3cm]{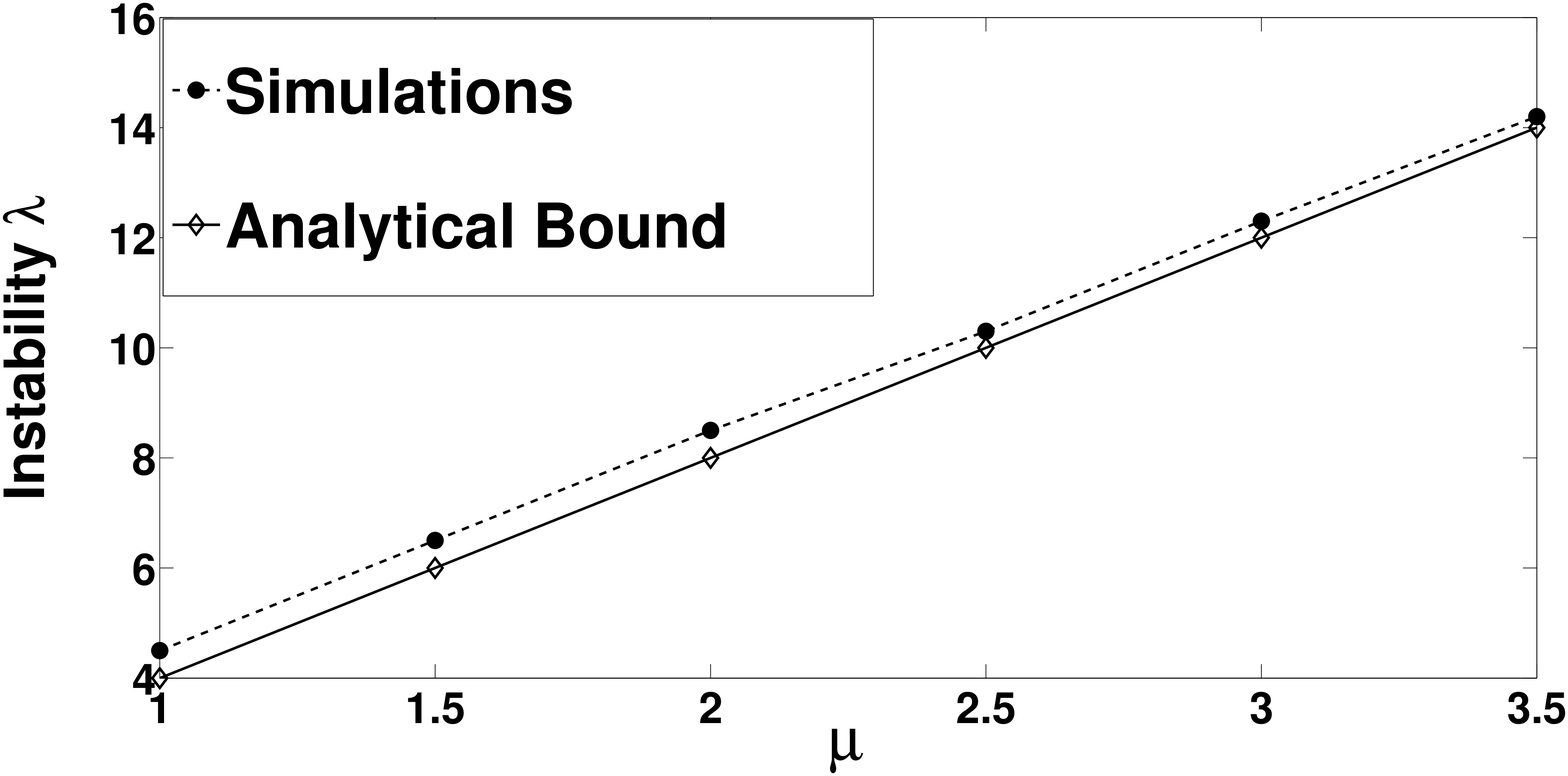}
\caption{\hspace*{-4.5cm}}\label{4a}
\end{subfigure}
\end{minipage}
\hspace{0.2mm}
\begin{minipage}[b]{0.484\columnwidth}
\begin{subfigure}[t]{1in}
\centering
\includegraphics[width=4.7cm, height=3cm]{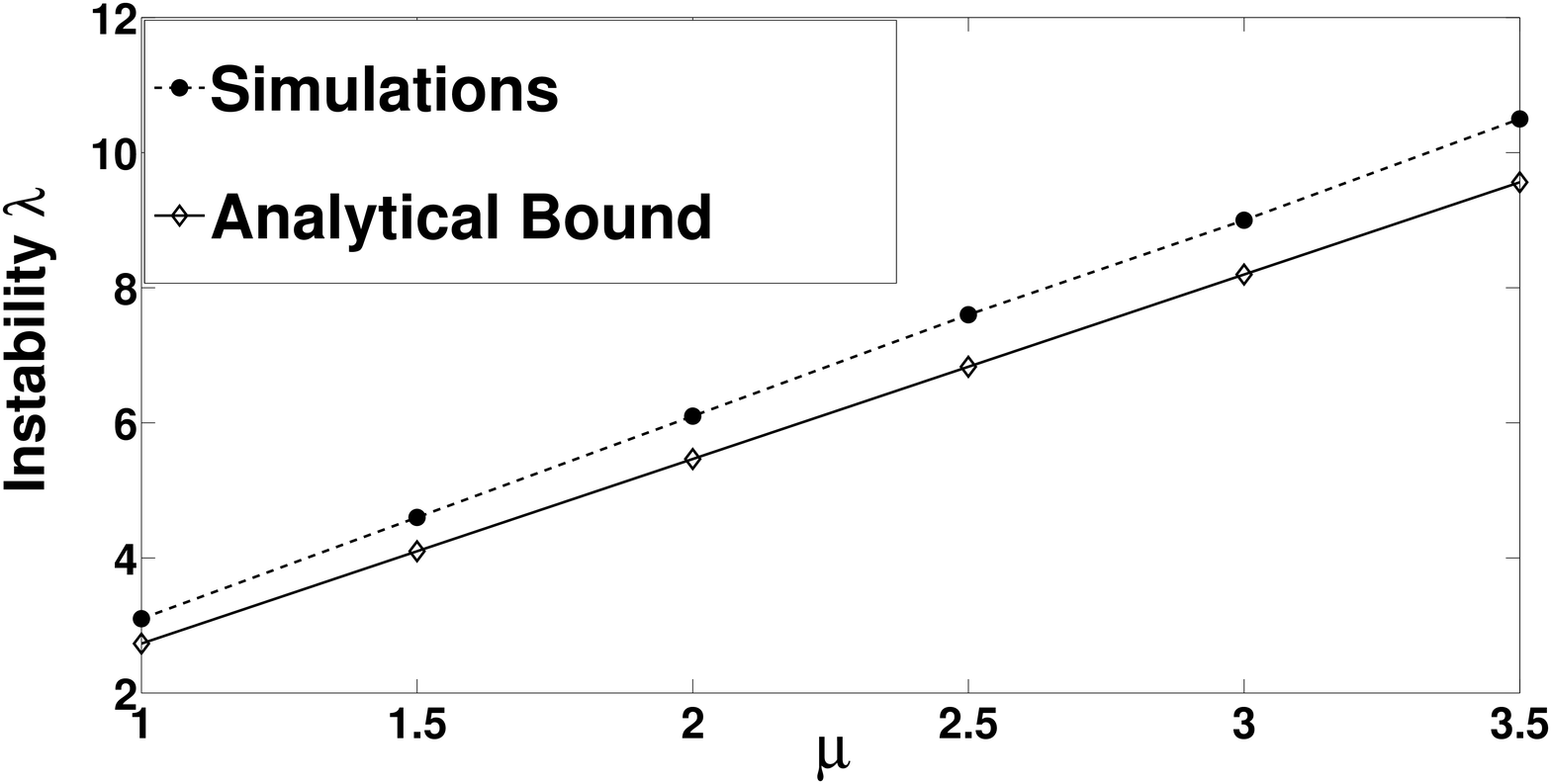}
\caption{\hspace*{-4.5cm}}\label{4b}
\end{subfigure}
\end{minipage}
\centering
\caption{The plot on the left (respectively, right) shows the minimum value of $\lambda$ for which instability is observed under the Mixed policy and the analytical bound shown on it in Theorem~\ref{thm:heuristic}(a) vs parameter $\mu$. The parameter values  $p_1 = 0.5$ (respectively, $p_1=1$) and $m_1 = m_2 = 1$ are used.}
\label{4}
\end{figure}

Fig.~\ref{gen}(\subref{gb}) (respectively,~\ref{3}(\subref{3a})) compares the instability threshold obtained via simulations and the bound $\frac{2 \mu \sum_{i=1}^{B}m_i}{2-\max(p_1, p_2, \ldots, p_B)}$ (respectively, $\frac{2\mu(m_1+m_2)}{1+\min(p_1,p_2)}$) shown in Theorem~\ref{load2}(a) (respectively, Theorem~\ref{thm:load}(a)) for the Load based policy in the arbitrary number of BSs case (respectively, two BSs case). Fig.~\ref{3}(\subref{3b}) compares the instability threshold obtained via simulations and the bound $3\mu$ shown in Theorem~\ref{thm:selfish}(a)  for the Throughput based policy. Fig.~\ref{4}(\subref{4a}) (respectively, \ref{4}(\subref{4b})) compares the instability threshold obtained via simulations and the bound $2\mu c_2$ (respectively, $(\sqrt{3}+1)\mu$) shown in Theorem~\ref{thm:heuristic}(a)  for the Mixed policy for $p_1= 0.5$ (respectively, $p_1= 1$). From these five plots, it can be seen that the bounds proved in Theorems~\ref{thm:load}(a),~\ref{thm:selfish}(a),~\ref{thm:heuristic}(a) and~\ref{load2}(a) are quite close to the actual instability thresholds.

\vspace{-.3cm}

\subsection{Simulation Results for Realistic Model}
\label{SSC:simulations:realistic}
Consider the model described in Section~\ref{model}. Suppose $B$ BSs are placed uniform at  random in a 1 $\mbox{km}^2$ area such that the distance between any two BSs is at least $d_{min}$, where $d_{min}$ is a parameter. Suppose the transmission power of each BS equals $P$, a constant. Let $N_0$ be the two sided noise power spectral density. The initial position of each arriving user is distributed uniformly at random in the square. We model the channels between BSs and users by considering path loss, lognormal shadowing, Rician fading and blocking~\cite{ric}. In particular,  the average received power at user $j$ from BS $i$ is proportional to $d_{ij}^{-\alpha}$, where $d_{ij}$ is the distance between BS $i$ and user $j$ and $\alpha$ is the path loss exponent.  We use the probabilistic model described in~\cite{los} to model  blocking. Specifically, the probability that the link between BS $i$ and user $j$ is LoS is:
\begin{equation}
P_{los}= e^{-d_{ij}/\beta}
\label{los}
\end{equation}
where $\beta = 200$ m. The path loss exponent of the link between BS $i$ and user $j$ depends on whether the link is LoS or NLoS and can be expressed as follows:
\begin{equation}
\alpha(d_{ij})=\left\{ 
\begin{array}{ll}
\alpha_l, & \mbox{with probability } P_{los}, \\
\alpha_n, & \mbox{with probability} \;P_{nlos}=1-P_{los},
\end{array}
\right.
\end{equation}
where $P_{los}$ is as in~\eqref{los}. The data rate $r_{ij}$ achieved at a user $j$ from a BS $i$ is given by the Shannon capacity of the channel from the BS $i$ to the user $j$. Users arrive into the network according to a Poisson process with parameter $\lambda$. Suppose the sizes of the files of different users are independent and  exponentially distributed random variables with parameter $\mu$.
A random way point (RWP) model~\cite{mmwn} has been adopted to model the mobility of the users present in the system. Specifically, at each point of a Poisson process with parameter $\lambda_m$,  a randomly chosen user moves to a different location, selected at random. In addition, at each point of another Poisson process with parameter $\lambda_c$, the quality of the link of a randomly chosen user to the BS it is associated with changes. 
Let the  $K$-factor values of Rician fading~\cite{ric} for LoS and NLoS links  be $K_l$ and $K_n$ respectively. Also, unless otherwise mentioned, the parameter $\theta$ in the Mixed policy (see \eqref{mixedeq}) is taken to be $0.2$. Table~\ref{t1} shows the value of various simulation parameters~\cite{parameter} for realistic model. 
\begin{table}
\centering
\caption{Simulation Parameters for Realistic Model}
\begin{tabular}{|c|c|}
\hline
Parameter & Value\\
\hline
P & \begin{tabular}{@{}c@{}}
46dBm\end{tabular}\\
$N_o$ & -90dBm/Hz\\
 $\mu$ & 100Gb\\
 $d_{min}$& 100m\\
$\alpha_l$  & 3.5 \\
$\alpha_n$ & 5.7\\
 $K_l$&  12dB  \\
$K_n$  & 5dB \\
Bandwidth & 250MHz\\
$\lambda_c$ & 3$s^{-1}$\\
$\lambda_m$& 1$s^{-1}$\\
 \hline
\end{tabular}
\label{t1}
\end{table} 
 
\begin{figure}[!hbt]
\begin{minipage}[b]{0.484\columnwidth}
\begin{subfigure}[t]{1in}
\centering
\includegraphics[width=4.7cm, height=3cm]{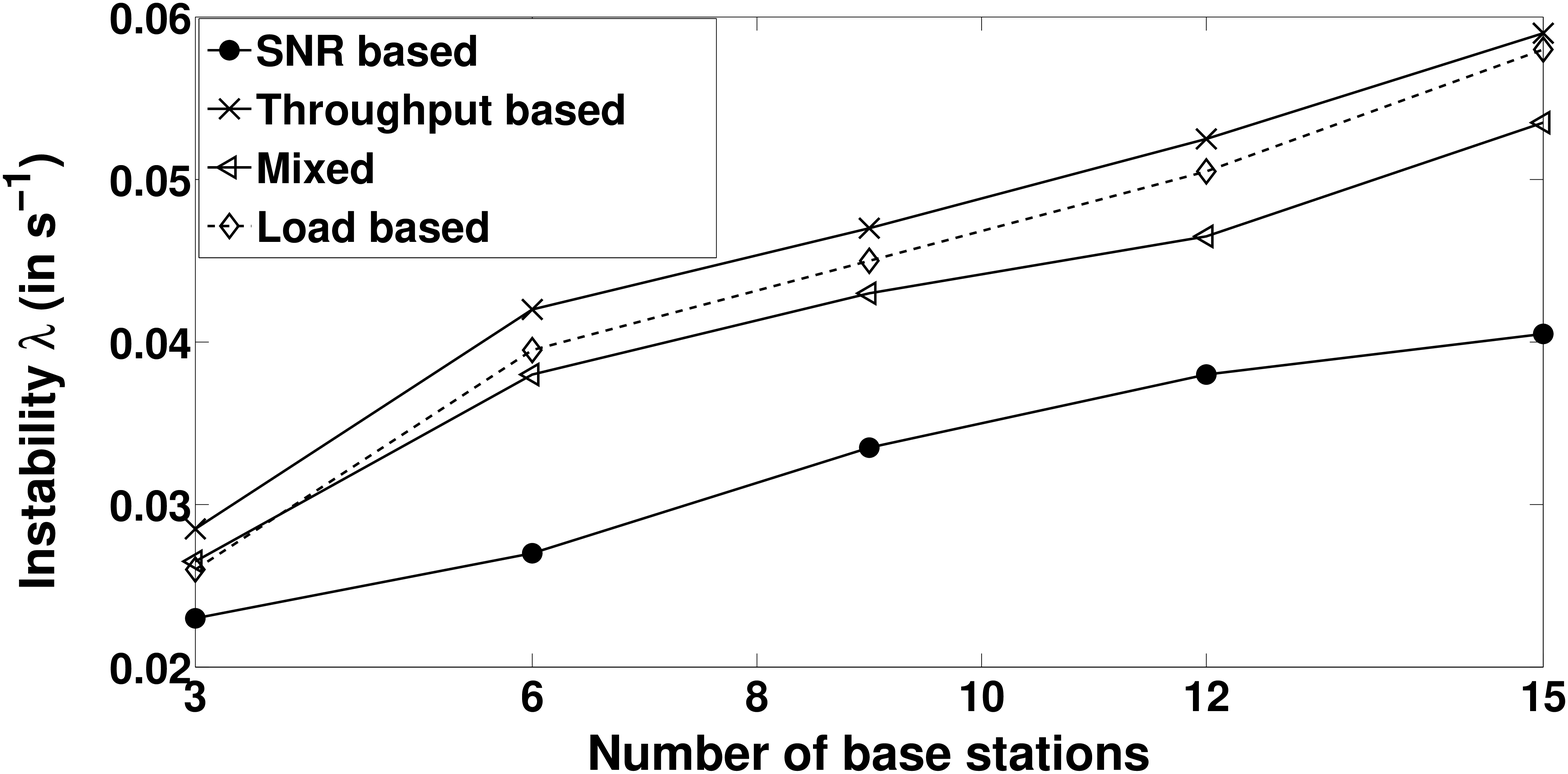}
\caption{\hspace*{-2.5cm}}\label{5a}
\end{subfigure}
\end{minipage}
\hspace{0.2mm}
\begin{minipage}[b]{0.484\columnwidth}
\begin{subfigure}[t]{1in}
\centering
\includegraphics[width=4.7cm, height=3cm]{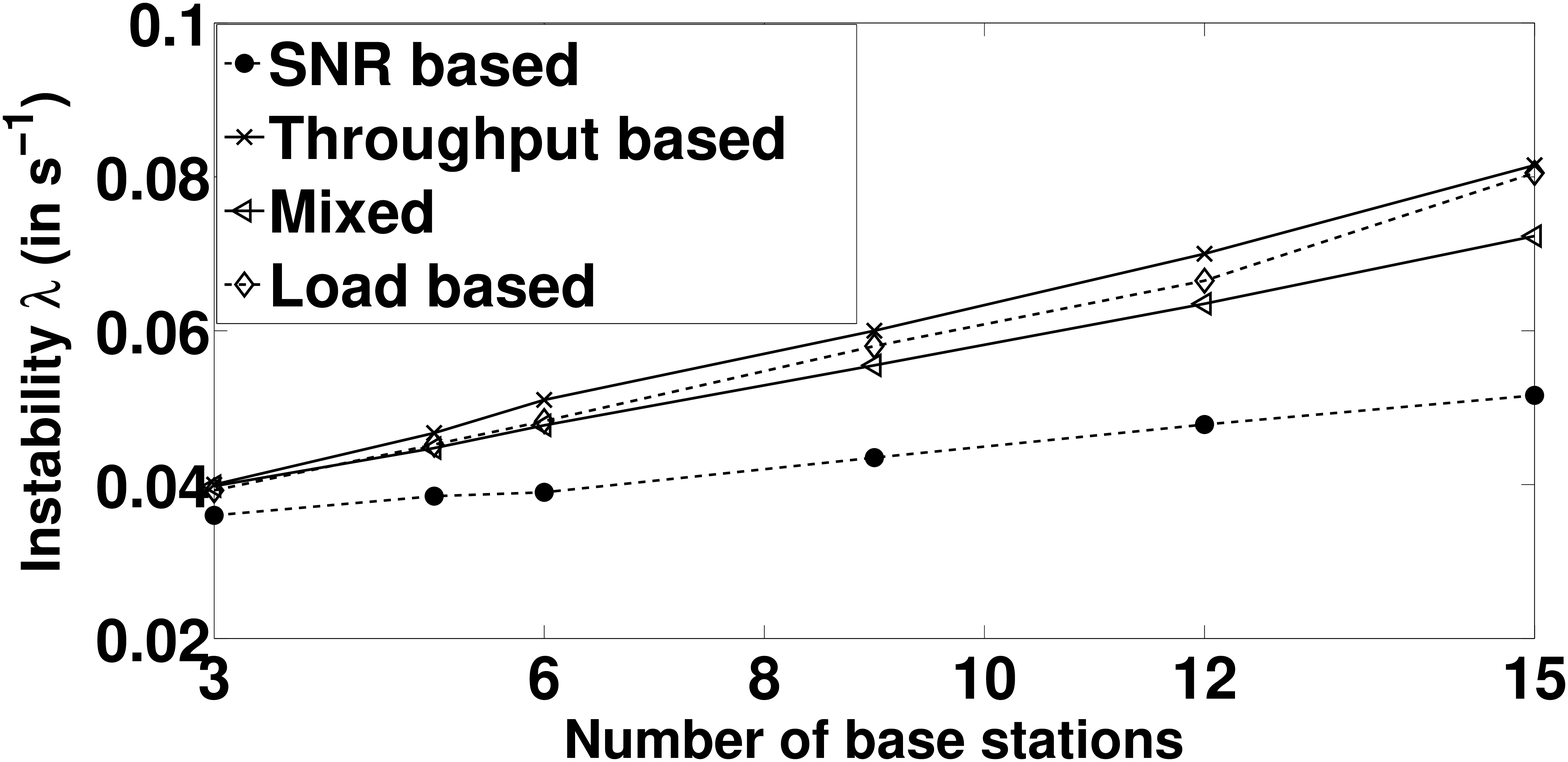}
\caption{\hspace*{-2.5cm}}\label{5b}
\end{subfigure}
\end{minipage}
\centering
\caption{The plot on the left (respectively, right) shows the minimum value of $\lambda$ at which instability is observed vs the number of base stations, $B$, for a heterogeneous system (respectively, homogeneous system with $3$ RF chains per BS).}
\label{5}
\end{figure}

\begin{figure}[!hbt]
\begin{minipage}[b]{0.484\columnwidth}
\begin{subfigure}[t]{1in}
\centering
\includegraphics[width=4.7cm, height=3cm]{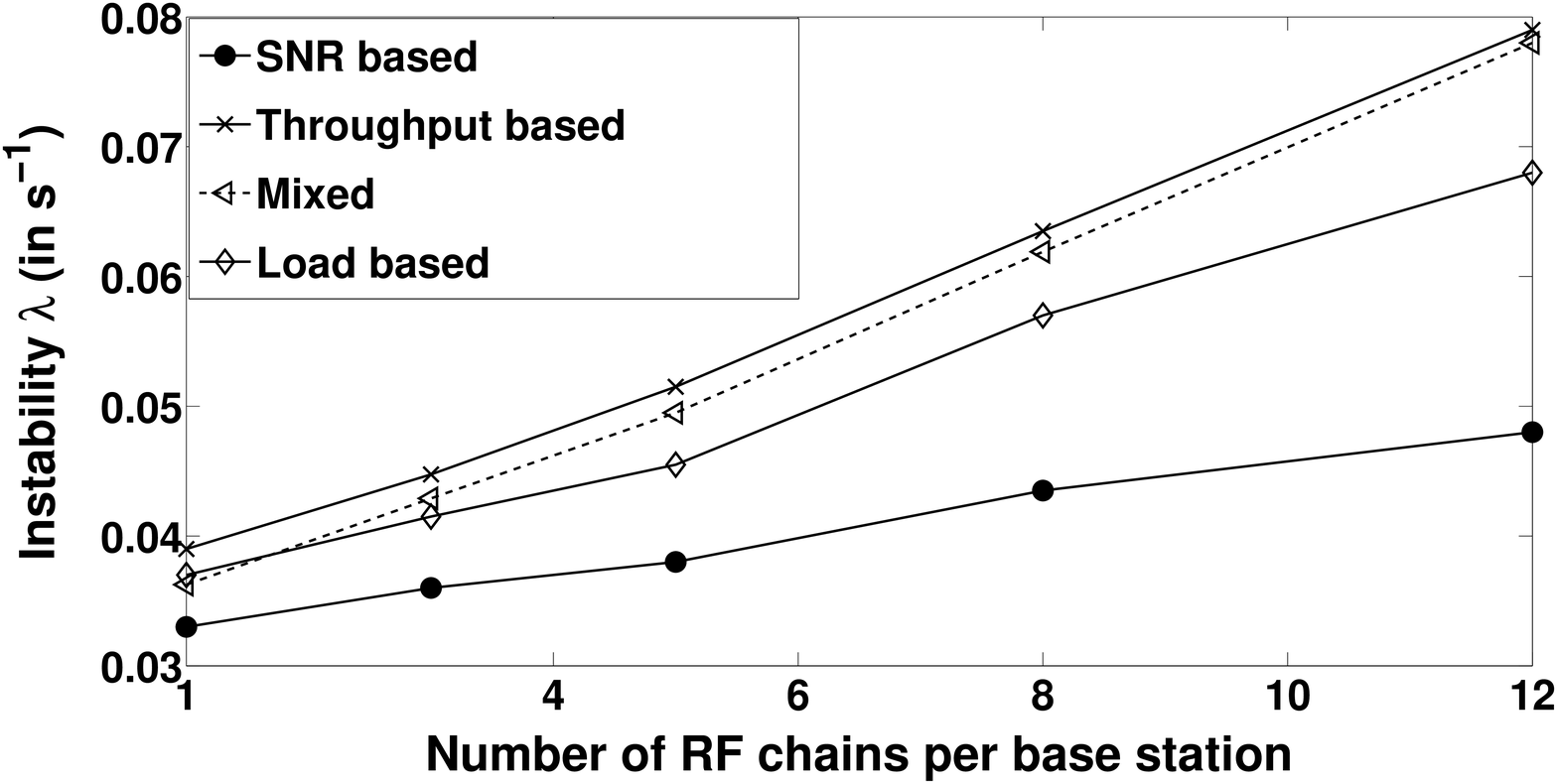}
\caption{\hspace*{-2.5cm}}\label{6a}
\end{subfigure}
\end{minipage}
\hspace{0.2mm}
\begin{minipage}[b]{0.484\columnwidth}
\begin{subfigure}[t]{1in}
\centering
\includegraphics[width=4.7cm, height=3cm]{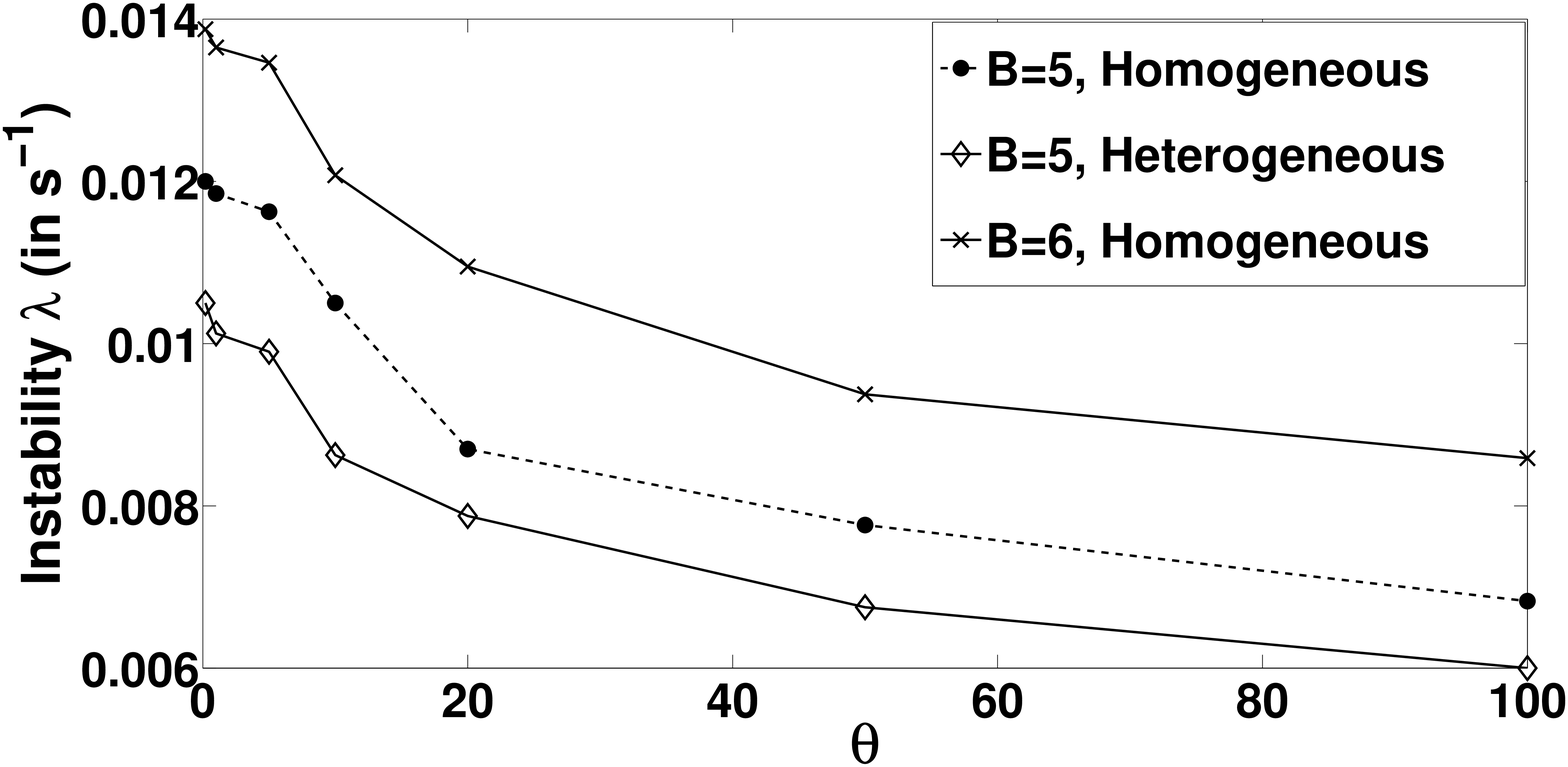}
\caption{\hspace*{-2.5cm}}\label{6b}
\end{subfigure}
\end{minipage}
\centering
\caption{The plot on the left (respectively, right) shows the minimum value of $\lambda$ at which instability is observed vs the number of RF chains per BS (respectively, parameter $\theta$ in the Mixed policy) for a homogeneous system with $B = 5$ BSs (respectively, for homogeneous systems with $B = 5$ BSs and $B = 6$ BSs and a heterogeneous system with $B = 5$ BSs).} 
\label{6}
\end{figure}

We consider two types of systems-- homogeneous and heterogeneous. In the homogeneous system, all BSs have equal numbers of RF chains. In the heterogeneous system, each of the odd numbered (respectively, even numbered) BSs has two (respectively, three) RF chains.  Fig.~\ref{5}(\subref{5a}) (respectively, Fig.~\ref{5}(\subref{5b})) shows the variation of the instability threshold, i.e., the minimum value of the user arrival rate, $\lambda$, at which the system becomes unstable, versus the number of BSs, $B$, in a heterogeneous (respectively, homogeneous) system. In both plots, the instability threshold  increases as the number of BSs increases, which is consistent with intuition-- more BSs can support a higher user load. For a homogeneous system, Fig.~\ref{6}(\subref{6a}) shows the variation of the instability threshold versus the number of RF chains each BS has. It can be seen that the instability threshold  increases as the number of RF chains increases; this is consistent with intuition since the greater the number of RF chains at a BS, the more the number of users that can be served simultaneously by the BS. Also, in all three plots (Figs.~\ref{5}(\subref{5a}),~\ref{5}(\subref{5b}) and~\ref{6}(\subref{6a})), the Throughput based policy outperforms the SNR based, Load based and Mixed policies and the performance of the SNR based policy is the worst among the four policies. 

Fig.~\ref{6}(\subref{6b}) shows the variation of the instability threshold with the parameter $\theta$ in the Mixed policy (see \eqref{mixedeq}). The instability threshold decreases monotonically with the parameter $\theta$. This is consistent with the trends observed in Figs.~\ref{5}(\subref{5a}),~\ref{5}(\subref{5b}) and~\ref{6}(\subref{6a}) (in which the Throughput based policy outperforms the Mixed policy with $\theta = 0.2$, which in turn outperforms the SNR based policy)  because as $\theta$ increases, the Mixed policy changes from the Throughput based policy ($\theta =0$) to the SNR based policy ($\theta =\infty$).

\begin{figure}[!hbt]
\begin{minipage}[b]{0.484\columnwidth}
\begin{subfigure}[t]{1in}
\centering
\includegraphics[width=4.7cm, height=3cm]{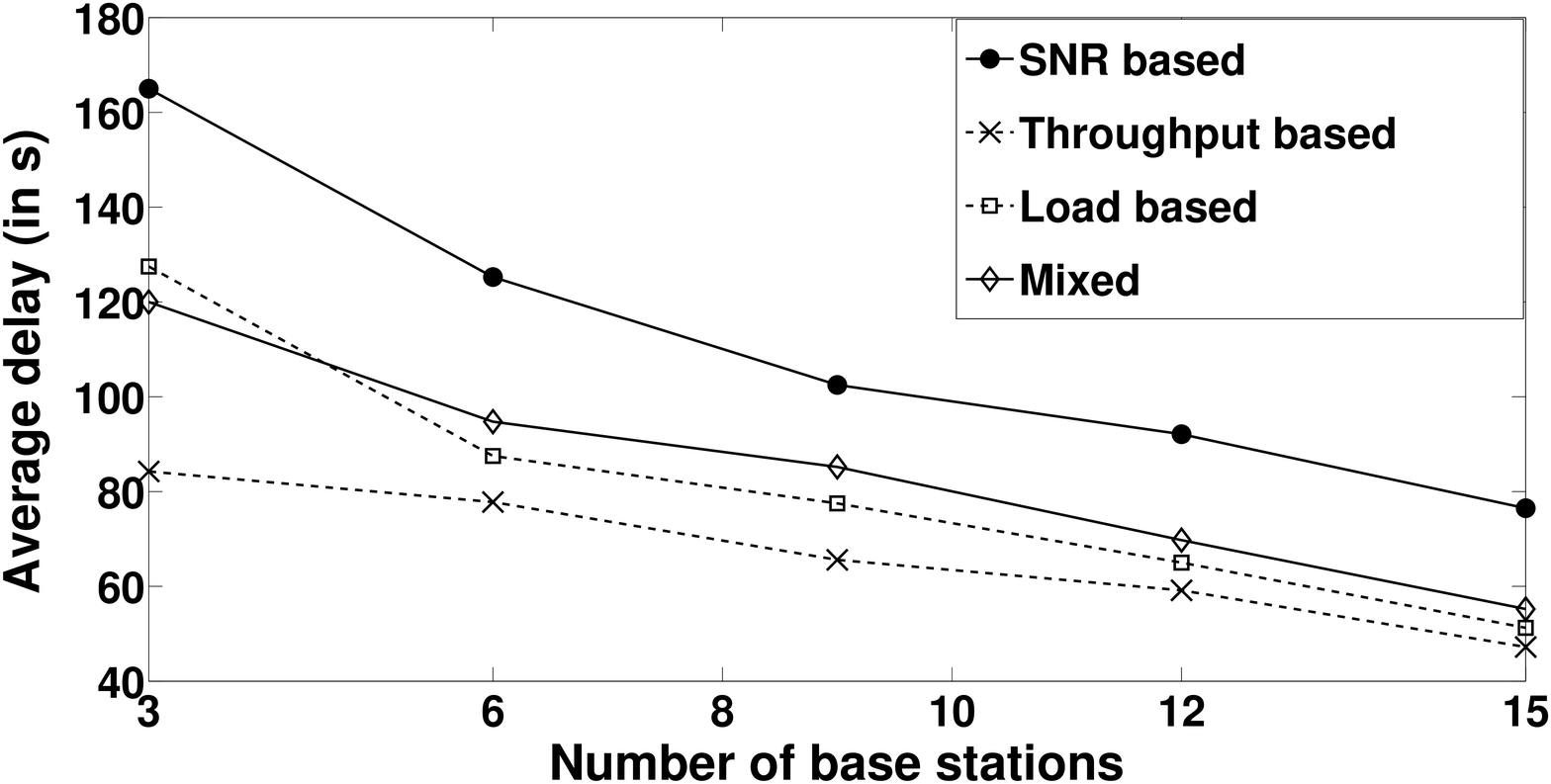}
\caption{\hspace*{-2.5cm}}\label{delay_a}
\end{subfigure}
\end{minipage}
\hspace{0.2mm}
\begin{minipage}[b]{0.484\columnwidth}
\begin{subfigure}[t]{1in}
\centering
\includegraphics[width=4.7cm, height=3cm]{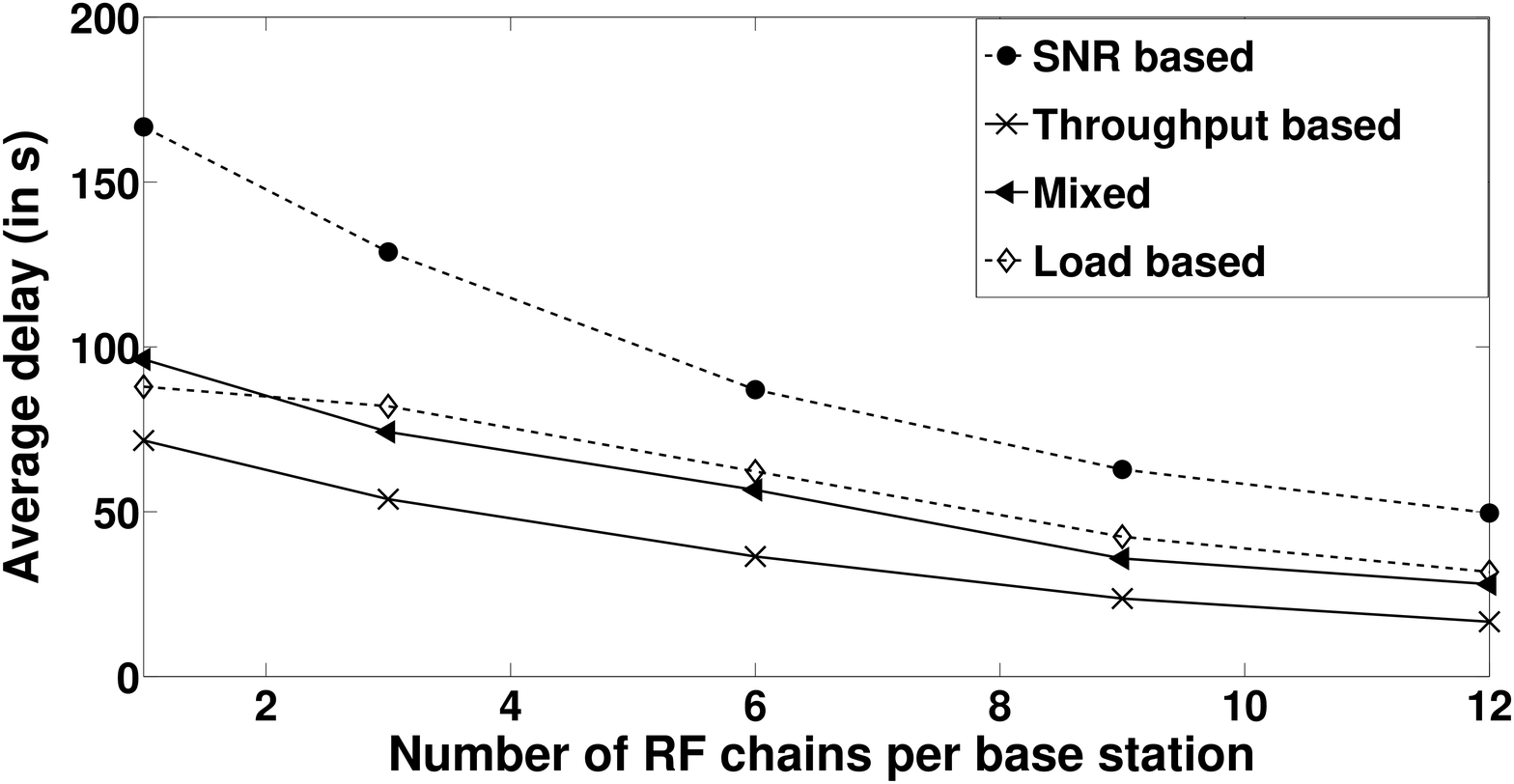}
\caption{\hspace*{-2.5cm}}\label{delay_b}
\end{subfigure}
\end{minipage}
\centering
\caption{The plot on the left (respectively, right) shows the average delay vs the number of BSs $B$ (respectively, the number of RF chains per BS) for $\lambda = 0.01 s^{-1}$ for a homogeneous system in which each BS has $3$ RF chains (respectively,  there are $B = 5$ BSs).}
\label{delay}
\end{figure}

Let the delay experienced by a user be the difference between the time when it departs from the system after transmitting its complete file and the time of its arrival into the system. Fig.~\ref{delay}(\subref{delay_a}) (respectively, ~\ref{delay}(\subref{delay_b})) shows the variation of the average delay per user with the number of BSs (respectively, the number of RF chains per BS)  for $\lambda = 0.01 \; s^{-1}$. Similar to the trends in Figs.~\ref{5}(\subref{5a}),~\ref{5}(\subref{5b}) and~\ref{6}(\subref{6a}), the average delay monotonically decreases as the number of BSs, $B$, or the number of RF chains per BS increases. Also, similar to the trends in those plots, the average delay is the least (respectively, most) under the Throughput based policy (respectively, SNR based policy). 

\begin{figure}[!hbt]
\begin{minipage}[b]{0.484\columnwidth}
\begin{subfigure}[t]{1in}
\centering
\includegraphics[width=4.7cm, height=3cm]{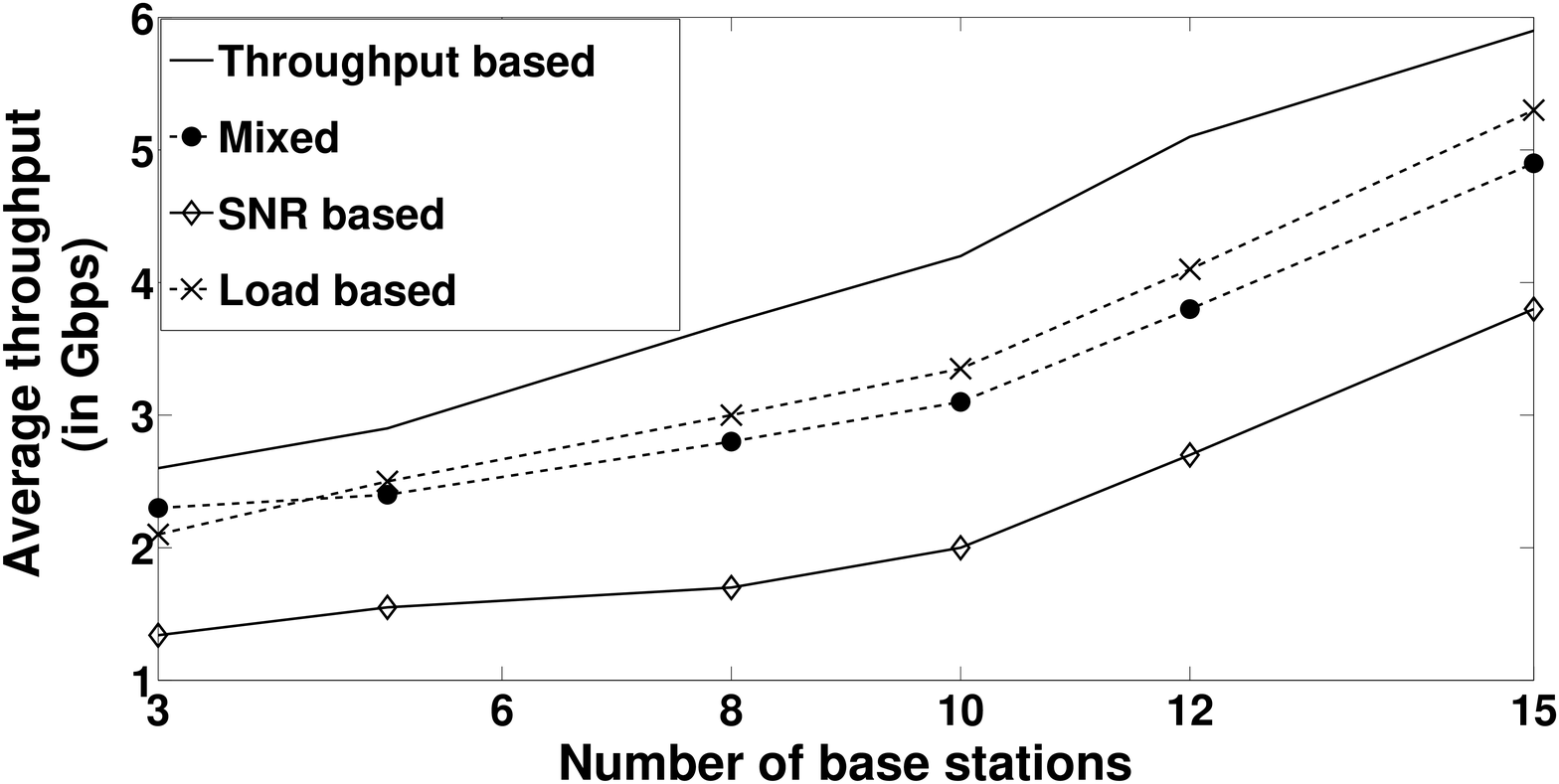}
\caption{\hspace*{-2.5cm}}\label{9a}
\end{subfigure}
\end{minipage}
\hspace{0.2mm}
\begin{minipage}[b]{0.484\columnwidth}
\begin{subfigure}[t]{1in}
\centering
\includegraphics[width=4.7cm, height=3cm]{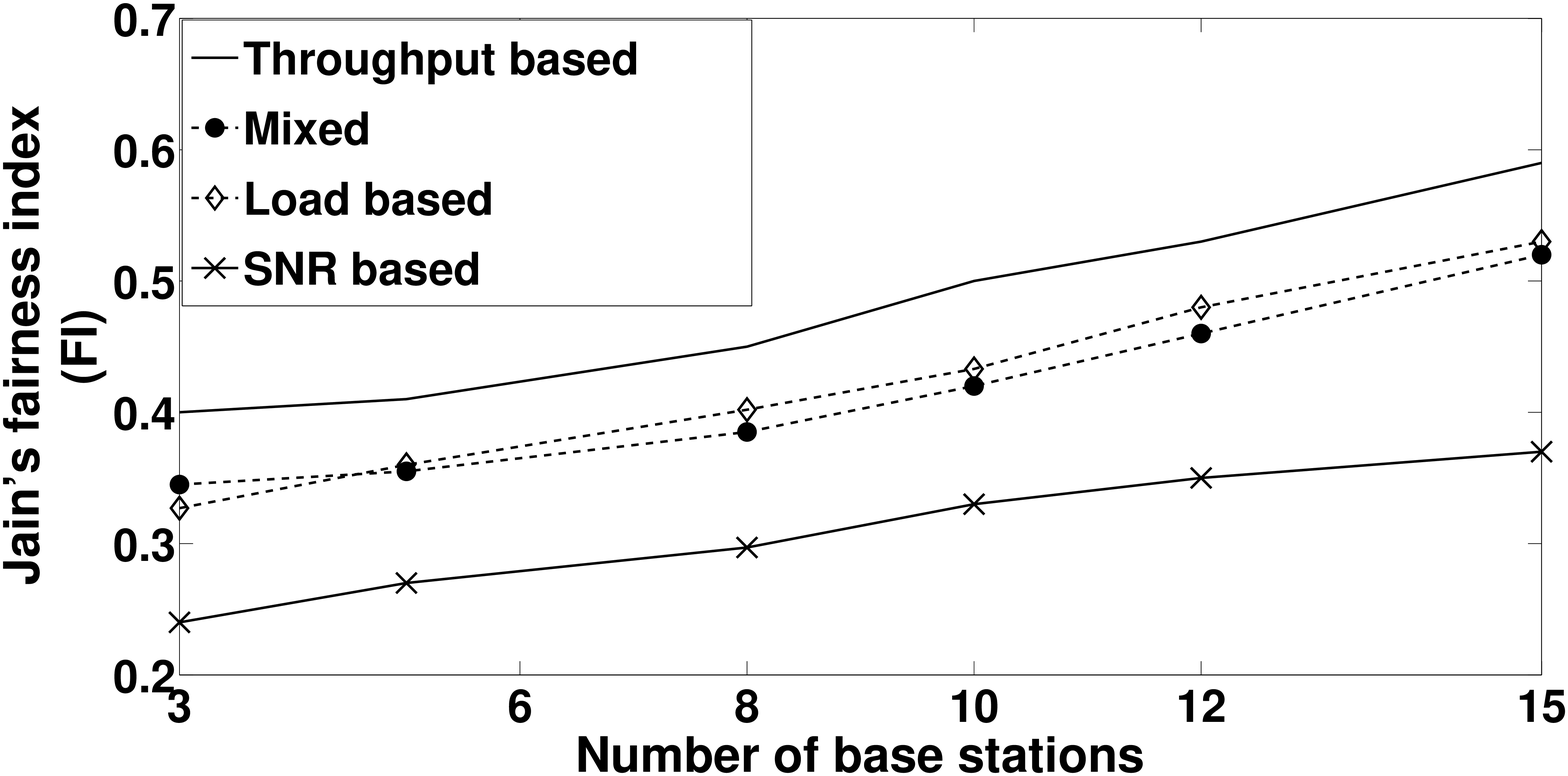}
\caption{\hspace*{-2.5cm}}\label{9b}
\end{subfigure}
\end{minipage}
\centering
\caption{The plot on the left (respectively, right) shows the average throughput (respectively, Jain's fairness index) vs the number of base stations, $B$, for $\lambda = 0.01 s^{-1}$ for a homogeneous system with $3$ RF chains per BS.}
\label{9}
\end{figure} 

Let $\mathcal{T}_i$, $\mathcal{F}_i$ and $\mathcal{D}_i$  be the average throughput, file size and delay respectively of the $i$'th user that departs from the system after transmitting its file. Note that  $\mathcal{T}_i=\frac{\mathcal{F}_i}{\mathcal{D}_i}$. Let $M$ be the total number of users who arrive into and depart from the system in a fixed and large time interval. We consider \emph{Jain's fairness index} as a fairness metric, which is defined as follows~\cite{Jain}: 
\begin{equation}
\label{fi}
\mbox{FI}= \frac{(\sum_{i=1}^{M} \mathcal{T}_i)^2}{M (\sum_{i=1}^{M} \mathcal{T}_i^{2})}.
\end{equation}
The value of FI lies between 0 and 1. Also, it increases with the degree of fairness of the distribution of average throughput; if all users get exactly equal average throughput, it takes value $1$ and it equals $\frac{q}{t}$ when exactly $q$ out of $t$ users have equal average throughput and the remaining $(t-q)$ users have $0$ average throughput~\cite{Jain}. See~\cite{Jain} for further properties of the fairness index. 
Fig.~\ref{9}(\subref{9a}) (respectively,~\ref{9}(\subref{9b})) shows the variation of average throughput (respectively, Jain's fairness index) with the number of BSs for $\lambda = 0.01 \; s^{-1}$. Similar to the trends in Figs.~\ref{5} and~\ref{delay}, the Throughput based policy (respectively, SNR based policy) performs the best (respectively, worst) in terms of the average throughput as well as Jain's fairness index.   

In summary, our simulations show that the Throughput based policy outperforms the other three user association policies in terms of stability region as well as average throughput, average delay and fairness performance. In particular, the Throughput based policy consistently outperforms the Mixed policy. This is surprising since under a model similar to that in this paper, but in the context of sub-6GHz 802.11 Wireless Local Area Networks (WLANs), the Mixed policy was found to outperform the Throughput based policy~\cite{gsk}. 

The above trends can be intuitively explained as follows. In the model in~\cite{gsk}, if $\mathcal{X}_i$ is the set of users associated with BS $i$ and $r_{ij}$ is the data rate of user $j$ associated with BS $i$, then the throughput that each user associated with BS $i$ gets is $\frac{1}{\sum\limits_{j \in \mathcal{X}_i}\frac{1}{r_{ij}}}$. Hence, even if \emph{one} user is associated with BS $i$ at a low data rate, then \emph{all} users associated with BS $i$ get low throughput. In the Mixed policy, the term $\theta r_{ij}$ in \eqref{mixedeq} discourages the association of users with BSs at low data rates. Hence, the Mixed policy outperforms the Throughput based policy under the model in~\cite{gsk}.  However, in the model in this paper, the throughput of a user $j$ associated with BS $i$ is given by \eqref{thr} and is  independent of the data rates of the other users associated with BS $i$. Hence, the Throughput based policy outperforms the Mixed policy under the model in this paper.

\section{Conclusions}
\label{conclusion} 
In this paper, we analytically characterized the stability regions of 
four user association policies-- SNR based, Throughput
based, Load based and Mixed-- using a CTMC model and Lyapunov function techniques. Our results show that the bounds we analytically proved on the instability thresholds of various user association policies are quite close to the actual instability thresholds obtained via simulations.  We also evaluated the performance of the above four user association policies in a large mmWave network, in which link qualities fluctuate with time and users are mobile, via detailed simulations. Our simulations show that the Throughput based policy outperforms the other three user association policies in terms of stability region as well as average throughput, average delay and fairness performance. In particular, the Throughput based policy consistently outperforms the Mixed policy. This is surprising since under a model similar to that in this paper, but in the context of sub-6GHz 802.11 Wireless Local Area Networks (WLANs), the Mixed policy was found to outperform the Throughput based policy~\cite{gsk}.

\appendix
\begin{IEEEproof}[Proof of Theorem~\ref{thm:capacity}]
The proof is similar to that of Theorem 3 in~\cite{gsk} and is omitted for brevity.
\end{IEEEproof}

\begin{IEEEproof}[\quad Proof of Theorem~\ref{thm:SNR}] 
Note that Theorem~\ref{thm:SNR} is a special case of Theorem~\ref{snr2}, whose proof is provided later in this Appendix.
\end{IEEEproof}

\begin{IEEEproof}[\quad Proof of Theorem~\ref{thm:load}]

{\bf Proof of part (a):}

We prove part (a) using Theorem~\ref{unstable}.
Let,\\
$V(s)=\left(\frac{X_{11}+X_{21}}{c_1}+\frac{X_{12}+X_{22}}{c_2}\right)$.\\
Define $\Delta\tilde{V}(s)=\Delta V(s)\nu(s)$, where $\nu(s)$ is the total rate with which transitions out of state $s$ occur.
Note that $\Delta\tilde{V}(s) \geq 0 \Leftrightarrow \Delta V(s) \geq 0.$ 
We will show that $\Delta \tilde{V}(s) \geq 0 \;\; \forall s \notin S_0$ for $\lambda \geq \frac{2\mu(m_1+m_2)}{1+\min(p_1,p_2)}$, where the finite set $S_0 = \{ \left( \begin{smallmatrix} 0&0\\ 0&0 \end{smallmatrix} \right)\}$. 

Recall that $X_{11}+X_{12}= X_1$ and  $X_{21}+X_{22}=X_2$. First, assume that $X_{11}, X_{12}, X_{21}$ and $X_{22}$ are all non zero. 

 \textbf{ Case 1:}  $m_1 \leq X_1$ and $m_2 \leq X_2$ 
 \begin{enumerate}[(I)]
 \item
 If $X_1 > X_2$, then an arriving user will associate with BS 2.
 \begin{eqnarray}
 \Delta \tilde{V}(s) &\hspace{-.6cm}= &\hspace{-.6cm}\lambda p_1\left(\frac{1}{c_1}\right)+ \lambda p_2 \left(\frac{1}{c_2} \right)+\frac{\mu X_{11} c_1 m_1}{X_{11}+X_{12}}\left(\frac{-1}{ c_1}\right) \nonumber\\
 & + &\hspace{-.2cm}\frac{\mu X_{12} c_2 m_1}{X_{11}+X_{12}}\left(\frac{-1}{ c_2}\right)+\frac{\mu X_{21} c_1 m_2}{X_{21}+X_{22}}\left(\frac{-1}{ c_1}\right) \nonumber \\
 & +&\hspace{-.2cm}\frac{\mu X_{22}c_2 m_2}{X_{21}+X_{22}}\left(\frac{-1}{ c_2}\right).  
 \end{eqnarray}
So $\Delta \tilde{V}(s)= \lambda\left(\frac{p_1}{c_1}+\frac{p_2}{c_2}\right)-\mu(m_1+m_2)$. Hence, using $c_1 = 1$, $c_2 = 2$ and $p_2 = 1 - p_1$, we get that $\Delta \tilde{V}(s) \geq 0$ iff $\lambda \geq\frac{2\mu(m_1+m_2)}{1+p_1}$.
\item 
If $X_2 > X_1$, then an arriving user will associate with BS 1.
 \begin{equation}
\Delta \tilde{V}(s) = \lambda p_1\left(\frac{1}{c_2}\right)+ \lambda p_2 \left(\frac{1}{c_1} \right)-\mu(m_1+m_2).
\end{equation}
So 
$\Delta \tilde{V}(s) = \lambda\left(\frac{p_1}{c_2}+\frac{p_2}{c_1}\right)-\mu(m_1+m_2)$. Hence, $\Delta \tilde{V}(s) \geq 0$ iff $\lambda \geq \frac{2\mu(m_1+m_2)}{1+p_2}$.
\item
If $X_2 = X_1$, then an arriving user will choose BS 1 and BS 2 with equal probabilities.
\begin{eqnarray}
\Delta \tilde{V}(s) & \hspace{-.7cm}= &\hspace{-.6cm}\frac{\lambda p_1}{2}\left(\frac{1}{c_2}+\frac{1}{c_1} \right)+ \frac{\lambda p_2}{2} \left(\frac{1}{c_1} +\frac{1}{c_2}\right) \nonumber \\
& \hspace{0.5cm}-& \mu(m_1+m_2).  
\end{eqnarray} 
\begin{equation}
\mbox{So}\,\, \Delta \tilde{V}(s)= \frac{\lambda}{2}\left(\frac{1}{c_2}+\frac{1}{c_1}\right)-\mu(m_1+m_2). 
\end{equation} 
$\Delta \tilde{V}(s) \geq 0$ iff $\lambda \geq \frac{2\mu(m_1+m_2)}{\frac{1}{c_2}+\frac{1}{c_1}} =\frac{2\mu(m_1+m_2)}{\frac{1}{2}+1}$. Since $\min(p_1,p_2) \leq \frac{1}{2}$, it follows that  $\Delta \tilde{V}(s) \geq 0$ if $\lambda \geq \frac{2\mu (m_1+m_2)}{1+\min(p_1,p_2)}$.
\end{enumerate}
\textbf{Case 2:}  $m_1 > X_1$ and $m_2 \leq X_2$ 
\begin{enumerate}[(I)]
\item
If $X_1 > X_2$, then an arriving user will associate with BS 2.
 \begin{eqnarray}
\Delta \tilde{V}(s) &\hspace{-0.5cm}= &\hspace{-0.5cm}\lambda p_1\left(\frac{1}{c_1}\right)+ \lambda p_2 \left(\frac{1}{c_2} \right)-\frac{\mu X_{21} m_2}{X_{21}+X_{22}}\nonumber \\
& \hspace{0.5cm}-&\frac{\mu X_{22} m_2}{X_{21}+X_{22}}-\frac{\mu X_{11} (X_{11}+X_{12})}{X_{11}+X_{12}}\nonumber \\
& \hspace{0.5cm}-&\frac{\mu X_{12}(X_{11}+X_{12})}{X_{11}+X_{12}}.
 \end{eqnarray}
So $\Delta \tilde{V}(s)= \lambda\left(\frac{p_1}{c_1}+\frac{p_2}{c_2}\right)-\mu (X_{11}+X_{12})-\mu m_2$. Since $m_1 > X_1 = X_{11} + X_{12}$, we get:\\
$\Delta \tilde{V}(s) \geq \lambda\left(\frac{p_1}{c_1}+\frac{p_2}{c_2}\right)-\mu m_1-\mu m_2 \geq 0$ 
iff $\lambda \geq\frac{2\mu(m_1+m_2)}{1+p_1}$. So $\Delta \tilde{V}(s) \geq 0$ if $\lambda \geq \frac{2\mu (m_1+m_2)}{1+\min(p_1,p_2)}$. 
\item
If $X_2 > X_1$, then an arriving user will associate with BS 1.
\begin{eqnarray}
\Delta \tilde{V}(s)&\hspace{-0.5cm}=& \hspace{-0.6cm}\lambda p_1\left(\frac{1}{c_2}\right)+ \lambda p_2 \left(\frac{1}{c_1} \right)-\frac{\mu X_{21}m_2}{X_{21}+X_{22}}\nonumber \\
& \hspace{0.5cm}-&\mu (X_{11}+X_{12})-\mu m_2.
\end{eqnarray} 
So $\Delta \tilde{V}(s)= \lambda\left(\frac{p_1}{c_2}+\frac{p_2}{c_1}\right)-\mu (X_{11}+X_{12})-\mu m_2.$\\
$\Delta \tilde{V}(s) \geq \lambda\left(\frac{p_1}{c_2}+\frac{p_2}{c_1}\right)-\mu m_1-\mu m_2 \geq 0$
iff $\lambda \geq \frac{2\mu(m_1+m_2)}{1+p_2}$. So $\Delta \tilde{V}(s) \geq 0$ if $\lambda \geq \frac{2\mu (m_1+m_2)}{1+\min(p_1,p_2)}$.
\item
If $X_2 = X_1$, then an arriving user will choose BS 1 and BS 2 with equal probabilities.
\begin{eqnarray}
\Delta \tilde{V}(s)& \hspace{-0.5cm}= &\hspace{-0.6cm}\frac{\lambda p_1}{2}\left(\frac{1}{c_2}+\frac{1}{c_1} \right)+ \frac{\lambda p_2}{2} \left(\frac{1}{c_1} +\frac{1}{c_2}\right)\nonumber \\
& \hspace{0.5cm}- &\mu (X_{11}+X_{12})-\mu m_2.
\end{eqnarray}
So  $\Delta \tilde{V}(s)= \frac{\lambda}{2}\left(\frac{1}{c_2}+\frac{1}{c_1}\right)-\mu (X_{11}+X_{12})-\mu m_2.$\\
$\Delta \tilde{V}(s) \geq \frac{\lambda}{2}\left(\frac{1}{c_2}+\frac{1}{c_1}\right)-\mu m_1-\mu m_2 \geq 0$\\
iff $\lambda \geq \frac{2\mu(m_1+m_2)}{\frac{1}{c_2}+\frac{1}{c_1}} =\frac{2\mu(m_1+m_2)}{\frac{1}{2}+1}$. Since $\min(p_1,p_2) \leq \frac{1}{2}$, $\Delta \tilde{V}(s) \geq 0$ if $\lambda \geq \frac{2\mu (m_1+m_2)}{1+\min(p_1,p_2)}$. 
\end{enumerate}
\textbf{Case 3:}  $m_1 \leq X_1$ and $m_2 > X_2$ \\
This case is symmetrical with Case (b) and hence we omit the details.

\textbf{Case 4:} $m_1 > X_1$ and $m_2 > X_2$ 
\begin{enumerate}[(I)]
\item
If $X_1 > X_2$, then an arriving user will associate with BS 2.
\begin{eqnarray}
\Delta \tilde{V}(s) &\hspace{-0.5cm}= &\hspace{-0.5cm}\lambda p_1\left(\frac{1}{c_1}\right)+ \lambda p_2 \left(\frac{1}{c_2} \right)\nonumber \\
& \hspace{0.5cm}- &\hspace{-0.3cm}\frac{\mu X_{11}(X_{11}+X_{12})}{X_{11}+X_{12}}-\frac{\mu X_{12} (X_{11}+X_{12})}{X_{11}+X_{12}}\nonumber \\
& \hspace{0.5cm}- &\hspace{-0.3cm}\frac{\mu X_{21}(X_{21}+X_{22})}{X_{21}+X_{22}}-\frac{\mu X_{22} (X_{21}+X_{22})}{X_{21}+X_{22}}. \nonumber
\end{eqnarray} 
\begin{equation}
\mbox{So}\,\, \Delta \tilde{V}(s)= \lambda\left(\frac{p_1}{c_1}+\frac{p_2}{c_2}\right)-\mu (X_{11}+X_{12}+X_{21}+X_{22}).
\end{equation}

Since $m_1 > X_1$ and $m_2 > X_2$,
$\Delta \tilde{V}(s) \geq \lambda\left(\frac{p_1}{c_1}+\frac{p_2}{c_2}\right)-\mu (m_1+ m_2) \geq 0$ 
iff $\lambda \geq\frac{2\mu(m_1+m_2)}{1+p_1}$. Since $\min(p_1,p_2) \leq p_1$, $\Delta \tilde{V}(s) \geq 0$ if $\lambda \geq \frac{2\mu (m_1+m_2)}{1+\min(p_1,p_2)}$. 
\item 
If $X_2 > X_1$, then an arriving user will associate with BS 1.
\begin{eqnarray}
\Delta \tilde{V}(s)&\hspace{-0.5cm}= &\hspace{-0.6cm}\lambda p_1\left(\frac{1}{c_2}\right)+ \lambda p_2 \left(\frac{1}{c_1} \right)\nonumber \\
& \hspace{0.5cm}-&\mu (X_{11}+X_{12}+X_{21}+X_{22}).
\end{eqnarray} 
\begin{equation}
\mbox{So}\,\, \Delta \tilde{V}(s)= \lambda\left(\frac{p_1}{c_2}+\frac{p_2}{c_1}\right)-\mu (X_{11}+X_{12}+X_{21}+X_{22}).
\end{equation}
$\Delta \tilde{V}(s) \geq \lambda\left(\frac{p_1}{c_2}+\frac{p_2}{c_1}\right)-\mu (m_1+ m_2) \geq 0$ 
iff $\lambda \geq\frac{2\mu(m_1+m_2)}{1+p_2}$. Hence, $\Delta \tilde{V}(s) \geq 0$ if $\lambda \geq \frac{2\mu (m_1+m_2)}{1+\min(p_1,p_2)}$.
\item
If $X_2 = X_1$, then an arriving user will choose BS 1 and BS 2 with equal probabilities.
\begin{eqnarray}
\Delta \tilde{V}(s)&\hspace{-0.5cm}= &\hspace{-0.6cm}\frac{\lambda p_1}{2}\left(\frac{1}{c_2}+\frac{1}{c_1} \right)+ \frac{\lambda p_2}{2} \left(\frac{1}{c_1} +\frac{1}{c_2}\right)
\nonumber \\
&\hspace{0.5cm} -& \mu (X_{11}+X_{12}+X_{21}+X_{22}).
\end{eqnarray} 
 \begin{equation}
\mbox{So}\,\, \Delta \tilde{V}(s)= \frac{\lambda}{2} \left(\frac{1}{c_2}+\frac{1}{c_1}\right)-\mu (X_{11}+X_{12}+X_{21}+X_{22}).
 \end{equation}

$\Delta \tilde{V}(s) \geq \frac{\lambda}{2} \left(\frac{1}{c_2}+\frac{1}{c_1}\right)-\mu (m_1+ m_2) \geq 0$\\
iff $\lambda \geq \frac{2\mu(m_1+m_2)}{\frac{1}{c_2}+\frac{1}{c_1}} =\frac{2\mu(m_1+m_2)}{\frac{1}{2}+1}$. Since $\min(p_1,p_2) \leq \frac{1}{2}$, $\Delta \tilde{V}(s) \geq 0$ if $\lambda \geq \frac{2\mu (m_1+m_2)}{1+\min(p_1,p_2)}$. 
\end{enumerate}

Now let us consider the case when some out of $X_{11}$, $X_{12}$, $X_{21}$, $X_{22}$ are zero. If ($X_{11} = 0, X_{12} \ne 0$ or $X_{12} = 0, X_{11} \ne 0$)
and/ or ($X_{21} = 0, X_{22} \ne 0$ or $X_{22} = 0, X_{21} \ne 0$), then in each of the above cases, the term(s) in the RHS of the expression for $\Delta\tilde{V}(s)$ corresponding to those $X_{ij}$, $\ i,j \in \{1,2\}$, which is/ are 0, disappear(s) and $\Delta\tilde{V}(s)\geq 0$ still holds. If $X_{11} = X_{12} = 0$ and
$X_{21} + X_{22} \ne 0$, then in each of the above cases, the  terms involving  $X_{11}$, $X_{12}$ disappear from the RHS of the expression for $\Delta\tilde{V}(s)$. Hence, $\Delta\tilde{V}(s) \ge 0$ still holds. Similarly, if $X_{21} = X_{22} = 0$ and
$X_{11} + X_{12} \ne 0$, then the  terms involving $X_{21}$, $X_{22}$ disappear  from the RHS of the expression for $\Delta\tilde{V}(s)$. Hence, $\Delta\tilde{V}(s) \ge 0$ still holds. 

Thus, in all possible cases, $\Delta\tilde{V}(s) \ge 0$ holds. Hence, condition \eqref{unstable1} in Theorem~\ref{unstable} holds. It is easy to check that all the other conditions in Theorem~\ref{unstable} hold. Hence, the system is unstable if $\lambda \geq \frac{2\mu (m_1+m_2)}{1+\min(p_1,p_2)}$. 

The result follows.

{\bf Proof of part (b):}
We now prove part (b) using Theorem~\ref{stable}. Let
$V(s)= X_{11}+X_{12}+X_{21}+X_{22}$ and define $\Delta\tilde{V}(s)=\Delta V(s)\nu(s)$ as in the proof of part (a). Let the finite set $S_0 = \{ \left( \begin{smallmatrix} 0&0\\ 0&0 \end{smallmatrix} \right)\}$. It is easy to show that $\nu(s)$ is lower and upper bounded by positive constants; hence, $\Delta\tilde{V}(s) \leq -\epsilon$ for some $\epsilon> 0$ and all $s \notin S_0$ iff $\Delta V(s) \leq -\epsilon^{\prime}$ for some $\epsilon^{\prime} > 0$ and all $s \notin S_0$. We will show that $\Delta \tilde{V}(s) \leq -\epsilon \;\; \forall s \notin S_0$ when $\lambda < \mu$, where $\epsilon = \mu - \lambda > 0 $.  Recall that $X_1= X_{11}+X_{12}$ and $X_2= X_{21}+X_{22}$.

Fix a state $s = \left( \begin{array}{cc}
X_{11} & X_{12} \\
X_{21} & X_{22}
\end{array} \right) \notin S_0$. 

{\bf Case 1}: $X_1 = 0$

In this case, an arriving user will associate with BS 1. 

$\Delta \tilde{V}(s)= \lambda p_1+ \lambda p_2+\frac{\mu X_{21} c_1 }{X_{21}+X_{22}}(-1)+\frac{\mu X_{22}c_2}{X_{21}+X_{22}}(-1)$.
  
So $\Delta \tilde{V}(s) = \lambda -\mu \frac{ X_{21} c_1+X_{22} c_2 }{X_{21}+X_{22}}$.
   
Since $c_2 > c_1$ and $c_1 = 1$, $\Delta \tilde{V}(s) \leq \lambda -\mu  c_1= \lambda - \mu = -\epsilon.$ 
    
{\bf Case 2}: $X_2 = 0$

In this case, an arriving user will associate with BS 2. This case is symmetrical with Case 1 and hence we omit the details.

{\bf Case 3}: $X_1 > X_2$ 

In this case, an arriving user will associate with BS 2.

First, assume that $X_2 > 0$. Then:
\begin{eqnarray}
\Delta \tilde{V}(s)&\hspace{-0.1cm}= \lambda p_1+ \lambda p_2+\frac{\mu X_{11} c_1}{X_{11}+X_{12}}(-1)+\frac{\mu X_{12} c_2}{X_{11}+X_{12}}(-1)\nonumber\\
& \hspace{-1.3cm}+ \frac{\mu X_{21} c_1}{X_{21}+X_{22}}(-1)+\frac{\mu X_{22}c_2}{X_{21}+X_{22}}(-1).
\end{eqnarray}  
So $\Delta \tilde{V}(s) = \lambda -\mu \frac{ X_{11} c_1+X_{12} c_2}{X_{11}+X_{12}}-\mu \frac{ X_{21} c_1+X_{22} c_2 }{X_{21}+X_{22}}$.
   
$\Delta \tilde{V}(s) \leq \lambda -\mu  c_1-\mu  c_1= \lambda - 2 \mu < -\epsilon$.

If $X_2 = 0$, then similar to the above, we get:
$\Delta \tilde{V}(s) = \lambda -\mu \frac{ X_{11} c_1+X_{12} c_2}{X_{11}+X_{12}} \leq \lambda -  \mu = -\epsilon.$

{\bf Case 4}: $X_1 < X_2$

In this case, an arriving user will associate with BS 1. This case is symmetrical with Case 3 and hence we omit the details.

{\bf Case 5}: $X_2 = X_1$

In this case, an arriving user will choose BS 1 and BS 2 with equal probabilities.
\begin{eqnarray}
\Delta \tilde{V}(s)&\hspace{-0.5cm}=&\hspace{-0.6cm} \lambda \frac{p_1}{2}+\lambda \frac{p_1}{2}+ \lambda \frac{p_2}{2}+\lambda \frac{p_2}{2}+\frac{\mu X_{11} c_1}{X_{11}+X_{12}}(-1)\nonumber\\
& \hspace{0.5cm}+&\frac{\mu X_{12} c_2}{X_{11}+X_{12}}(-1)+\frac{\mu X_{21} c_1}{X_{21}+X_{22}}(-1)\nonumber\\
& \hspace{0.5cm}+&\frac{\mu X_{22}c_2}{X_{21}+X_{22}}(-1).
\end{eqnarray} 
So $\Delta \tilde{V}(s) = \lambda -\mu \frac{ X_{11} c_1+X_{12} c_2}{X_{11}+X_{12}}-\mu \frac{ X_{21} c_1+X_{22} c_2 }{X_{21}+X_{22}}$.\\
   
$\Delta \tilde{V}(s) \leq \lambda -\mu  c_1-\mu  c_1= \lambda - 2\mu < -\epsilon$.

Thus, in all the cases, $\Delta \tilde{V}(s) \leq -\epsilon$. Hence, condition \eqref{stable3} in Theorem~\ref{stable} is satisfied if $\lambda  < \mu$. It is easy to check that the other conditions in Theorem~\ref{stable} are also satisfied. 

The result follows.
\end{IEEEproof}

\begin{IEEEproof}[\quad Proof of Theorem~\ref{thm:selfish}]
The proof of part (a) is similar to that of Theorem 5 in~\cite{gsk} and is omitted for brevity.

{\bf Proof of part (b):} We prove the result using Theorem~\ref{stable}. Since $p_1 = 1$, the rate vector of every arriving user is  $(c_2,c_1)$. Hence
$X_{11} = X_{22} = 0$. 
The Throughput based policy induces a 
CTMC with state $s = (X_{12} \, \, X_{21})$. 
Consider the Lyapunov function:
\begin{equation}
V(s) = X_{12}^2 + X_{21}^2 + X_{12} X_{21}.
\end{equation}
Let $\Delta\tilde{V} (s) =\Delta V(s) \nu(s)$, where $\nu(s)$ is the total rate at which transitions out of state $s$ occur. Let $\epsilon > 0$ be any positive number. 
Let $X_{\epsilon}$ be a large number. The particular conditions that 
 $X_{\epsilon}$ must satisfy will be stated later. 
Consider the sets of states $A = \{s: X_{12} > X_{\epsilon}\}$ and
$B = \{s: X_{21} > X_{\epsilon}\}$. 
We will show that  $\Delta \tilde{V} (s) \le - \epsilon$ for all states $s \in A \cup B$. The complement of $A \cup B$ is the finite set $S_0$ in Theorem~\ref{stable}. 

We show that condition \eqref{stable3} in Theorem~\ref{stable} holds. Fix a state $s \notin S_0$. Recall that under the Throughput based policy, an arriving user chooses the BS which will give it the higher throughput. That is, it chooses BS 1 if $\min(m_1,X_{12}+1)\frac{c_2}{X_{12}+1} > \min(m_2,X_{21}+1)\frac{c_1}{X_{21}+1}$, i.e., if  $\frac{c_2 m_1}{X_{12}+1} > \min(m_2,X_{21}+1)\frac{c_1}{X_{21}+1}$.  Consider the following two cases: 

\textbf{Case 1:} If $X_{21} \geq 1$, then since $\min(m_2,X_{21}+1) = m_2$, an arriving user chooses BS 1 if $\frac{c_2 m_1}{X_{12}+1} > \frac{c_1 m_2}{X_{21}+1}$,
i.e., if $X_{12} <  X_{21} $, chooses BS 2 if $X_{12} >  X_{21}$ and randomizes equally between BS 1 and BS 2 if $X_{12} =  X_{21}$.

\textbf{Case 2:} If $X_{21} =0$, then an arriving user chooses BS 1 if $\frac{c_2 m_1}{X_{12}+1} > (X_{21}+1)\frac{c_1}{X_{21}+1}$, i.e., if $X_{12} < 1$. However, note that the state $s = (X_{12} \, \, X_{21})$, where $X_{12} = X_{21} = 0$, is in $S_0$. Since  $s \notin S_0$, $X_{12} \neq 0$.  Therefore, if $X_{21} =0$, then an arriving user  always chooses BS 2. 

We now consider Cases 1 and 2 in detail:

\textbf{Case 1:}
We must have one of the following cases for state $s$:\\

\begin{enumerate}[(I)]
\item
($X_{12} > X_\epsilon, \, X_{21} \ne 0, \, X_{12} >  X_{21} )$: Then an arriving user associates with BS 2. Therefore, $\Delta \tilde{V} (s) = $\\
\begin{eqnarray}
\lambda \left\{ \left[ (X_{21}+1)^2 - X_{21}^2 \right] 
                     +  X_{12} (X_{21} + 1-X_{21}) \right\} \nonumber\\
& \hspace{-7.5cm}+  \mu c_2 m_1 \left\{ \left[ (X_{12}-1)^2 - X_{12}^2 \right]  
       + X_{21} (X_{12} - 1 - X_{12}) \right\}  \nonumber\\
& \hspace{-8.2cm}+  \mu c_1 \min(m_2,X_{21}) \left\{ \left[ (X_{21}-1)^2 - X_{21}^2 \right]  
       -  X_{12} \right\}. \nonumber
\end{eqnarray}
Consider the following subcases:\\
\textbf{(i)} $X_{21} =1$ and $\min(m_2,X_{21})=1$\\
\begin{equation}
\Delta \tilde{V} (s) = X_{12} (\lambda - 5 \mu) + \left( 3 \lambda
-   \mu \right).
\end{equation}
The first term on the RHS is negative for $\lambda < 4 \mu$. Hence 
$\Delta \tilde{V} (s) < -\epsilon$ for $X_\epsilon$ (and hence $X_{12}$) large enough.\\
\textbf{(ii)} $X_{21} >1$\\
\begin{equation}
\label{selfish:p1:case1}
\Delta \tilde{V} (s) = X_{12} (\lambda - 6 \mu) + 2X_{21} (\lambda - 3 \mu)+(\lambda + 4 \mu) .
\end{equation}
Consider the following two subcases:
\begin{itemize}
\item
$\lambda < 3 \mu$: The third term in the RHS of (\ref{selfish:p1:case1}) is constant. The first and second terms are negative. Thus $\Delta \tilde{V} (s) < - \epsilon$ for $X_\epsilon$ (and hence $X_{12}$) large enough. 
\item 
$\lambda \ge 3\mu$: $ X_{21} < X_{12} $ by assumption. So by (\ref{selfish:p1:case1}):
\begin{eqnarray}
\hspace{-.3cm}\Delta \tilde{V} (s)&\hspace{-.3cm}\leq &\hspace{-.1cm}X_{12} (\lambda - 6 \mu) + 2X_{12} (\lambda - 3 \mu)+(\lambda + 4 \mu) \nonumber\\
& \hspace{-.3cm}= & \hspace{-.1cm} 3X_{12}(\lambda-4\mu)+ (\lambda + 4 \mu).
 \label{EQ:Case1:b:b}
\end{eqnarray}
The first term on the RHS  of \eqref{EQ:Case1:b:b} is negative for $\lambda < 4 \mu$ and the second term is constant. Hence 
$\Delta \tilde{V} (s) < -\epsilon$ for $X_\epsilon$ (and hence $X_{12}$) large enough. 
\end{itemize}

\item
($X_{21} > X_\epsilon,  \, X_{12} < X_{21}, \, X_{12} \ne 0)$: Then, an arriving user associates with BS 1. Therefore, $\Delta \tilde{V} (s)  = $
\begin{eqnarray}
\lambda \left[ (X_{12}+1)^2 - X_{12}^2  + X_{21} (X_{12} + 1-X_{12}) \right] \nonumber\\
&  \hspace{-7.2cm}+ \mu c_2 m_1 \left[ (X_{12}-1)^2 - X_{12}^2 + X_{21} (X_{12} - 1 - X_{12}) \right]  \nonumber\\
&  \hspace{-9.2cm}+ \mu c_1 m_2 \left\{  \left[ (X_{21}-1)^2 - X_{21}^2 \right] 
     -  X_{12} \right\}.  
\end{eqnarray}
Substitution of values and algebraic simplification yields:
\begin{eqnarray}
\Delta \tilde{V} (s) & = & X_{12} (2 \lambda - 6 \mu) + X_{21} (\lambda - 6 \mu) \nonumber \\
                     &   &  + \left( \lambda + 4 \mu \right). \label{selfish:p1:case3}
\end{eqnarray}

Consider the following subcases:
\begin{itemize}
\item
$\lambda < 3 \mu$: The third term in the RHS of (\ref{selfish:p1:case3}) is constant. The first and second terms are negative. Thus $\Delta \tilde{V} (s) < - \epsilon$ for $X_\epsilon$ (and hence $X_{21}$) large enough. 
\item 
$\lambda \ge 3 \mu$: $X_{12} < X_{21}$ by assumption. So by (\ref{selfish:p1:case3}):
\begin{equation}
\Delta \tilde{V} (s) \le (3 \lambda - 12 \mu) X_{21}  + \lambda +5 \mu. 
\end{equation}
The first term on the RHS is negative for $\lambda <4 \mu$ and the second term is constant. Hence           
$\Delta \tilde{V} (s) < -\epsilon$ for $X_\epsilon$ (and hence $X_{21}$) large enough. 
\end{itemize} 

\item
$X_{21} > X_\epsilon$, $X_{12} = 0$: Then an arriving user associates with BS 1. So\; $\Delta \tilde{V} (s)$ 
\begin{eqnarray}
 & = & \lambda [1+X_{21}] + \mu c_1 m_2 \left\{\left[ (X_{21}-1)^2 - X_{21}^2  \right] \right\}. \nonumber \\
& = & \; X_{21}(\lambda - 4 \mu) + \left( \lambda + 2 \mu \right).
\end{eqnarray}
The first term is negative for $\lambda < 4 \mu$ and the second term is constant. Hence           
$\Delta \tilde{V} (s) < -\epsilon$ for $X_\epsilon$ (and hence $X_{21}$) large enough. 

\item
$X_{21} > X_\epsilon$, $X_{12} =  X_{21}$: Then an arriving user associates with BS 1 with probability $\frac{1}{2}$
and with BS 2 with probability $\frac{1}{2}$. Therefore, $\Delta \tilde{V} (s) =$ 
\begin{eqnarray} 
\hspace{.1cm}\frac{\lambda}{2} \left[ (X_{12}+1)^2 - X_{12}^2  + X_{21} (X_{12} + 1-X_{12}) \right]  \nonumber\\
& \hspace{-7.5cm}+ \frac{\lambda}{2} \left\{  \left[ (X_{21}+1)^2 - X_{21}^2 \right] + X_{12} (X_{21} + 1-X_{21}) \right\} \nonumber\\   
& \hspace{-7.2cm}+ \mu c_2 m_1 \left[ (X_{12}-1)^2 - X_{12}^2  + X_{21} (X_{12} - 1 - X_{12}) \right]  \nonumber\\
& \hspace{-9.2cm}+ \mu c_1 m_2 \left\{ \left[ (X_{21}-1)^2 - X_{21}^2 \right] - X_{12} \right\}. 
\end{eqnarray}
Simplifying and substituting $X_{21} = X_{12}$, $c_1 = 1, m_1 = 1$, $c_2 = 2,  m_2 = 2$ yields:
\begin{equation}
\Delta \tilde{V} (s) = X_{12}(3\lambda - 12 \mu) + \left( \lambda + 4 \mu \right).
\end{equation}
The first term on the RHS is negative for $\lambda < 4 \mu$ and the second term is constant. Hence           
$\Delta \tilde{V} (s) < -\epsilon$ for $X_\epsilon$ (and hence $X_{12}$) large enough. 
\end{enumerate}

\textbf{Case 2:} 
$X_{12} > X_\epsilon$, $X_{21} = 0$: Then, an arriving user associates with BS 2. 
\begin{eqnarray}
\Delta \tilde{V} (s) & = & \lambda \left[ 1 + X_{12} \right] + \mu c_2 m_1 [(X_{12}-1)^2 - X_{12}^2]
\nonumber \\
                     & = & X_{12} (\lambda - 4 \mu) + \left( \lambda + 2 \mu \right).
\end{eqnarray}
The first term on the RHS is negative for $\lambda < 4 \mu$ and the second term is constant. Hence           
$\Delta \tilde{V} (s) < -\epsilon$ for $X_\epsilon$ (and hence $X_{12}$) large enough.

Thus, in all the cases, $\Delta \tilde{V}(s) \leq -\epsilon$. Hence, condition \eqref{stable3} in Theorem~\ref{stable} is satisfied. It is easy to check that the other conditions in Theorem~\ref{stable} are also satisfied. 

The result follows.

{\bf Proof of part (c):} We prove the result using Theorem~\ref{stable}. For a state $s = \left( \begin{array}{cc}
X_{11} & X_{12} \\
X_{21} & X_{22}
\end{array} \right)$,  let
$S_1(s)=\left(\frac{X_{11}+X_{12}+1}{c_1}\right),\; S_2(s)= \left(\frac{X_{21}+X_{22}+1}{c_2}\right)$ and $V(s)=\left(\frac{X_{11}}{c_1}+\frac{X_{12}}{c_2}+\frac{X_{21}}{c_1}+\frac{X_{22}}{c_2}\right).$
We will show that $\Delta \tilde{V}(s) \leq -\epsilon \ \forall s  \notin S_0$ for $\lambda < \frac{4\mu}{3}$, where $\epsilon= \frac{1}{2}(\frac{4\mu}{3} -\lambda) > 0$ and the finite set $S_0 = \{ \left( \begin{smallmatrix} 0&0\\ 0&0 \end{smallmatrix} \right)\}$.
 
Consider a new user $j$ with rate vector $(c_2,c_1)$. Its throughput if it associates with BS 1 (respectively, 2) is $\mathcal{T}_{1j}= \frac{c_2 m_1}{X_{11}+X_{12}+1}$ (respectively,  $\mathcal{T}_{2j}= \frac{c_1 m_2}{X_{21}+X_{22}+1}$). So:
\begin{eqnarray}
\mathcal{T}_{1j}-\mathcal{T}_{2j}& = &\frac{\left(\frac{X_{21}+X_{22}+1}{c_1}-\frac{X_{11}+X_{12}+1}{c_2}\right)}{\left(\frac{X_{11}+X_{12}}{c_2}+\frac{1}{c_2}\right)\left(\frac{X_{21}+X_{22}}{c_1}+\frac{1}{c_1}\right)}\nonumber\\
 & =& \frac{2S_2-\frac{S_1}{2}}{\left(\frac{X_{11}+X_{12}}{c_2}+\frac{1}{c_2}\right)\left(\frac{X_{21}+X_{22}}{c_1}+\frac{1}{c_1}\right)}. 
\label{EQ:t1j:minus:t2j:c2:c1}
\end{eqnarray}

Consider a new user $j$ with rate vector $(c_1,c_2)$. Its throughput if it associates with BS 1 (respectively, 2) is $\mathcal{T}_{1j}= \frac{m_1 c_1}{X_{11}+X_{12}+1}$ (respectively, $\mathcal{T}_{2j}= \frac{m_2 c_2}{X_{21}+X_{22}+1}$). So:
\begin{eqnarray}
\mathcal{T}_{1j}-\mathcal{T}_{2j}& = &\frac{\left(\frac{X_{21}+X_{22}+1}{c_2}-\frac{X_{11}+X_{12}+1}{c_1}\right)}{\left(\frac{X_{11}+X_{12}}{c_1}+\frac{1}{c_1}\right)\left(\frac{X_{21}+X_{22}}{c_2}+\frac{1}{c_2}\right)}\nonumber\\
& = &\frac{S_2-S_1}{\left(\frac{X_{11}+X_{12}}{c_1}+\frac{1}{c_1}\right)\left(\frac{X_{21}+X_{22}}{c_2}+\frac{1}{c_2}\right)}.
\label{EQ:t1j:minus:t2j:c1:c2}
\end{eqnarray}

Consider the following cases:\\
\textbf{Case 1: $S_1=S_2$} \\
If a new user $j$ with rate vector $(c_2,c_1)$ arrives, it chooses BS 1 (because $\mathcal{T}_{1j}-\mathcal{T}_{2j} >0$ by \eqref{EQ:t1j:minus:t2j:c2:c1}). If a new user $j$ with rate vector $(c_1,c_2)$ arrives, the throughput that it will get with the two BSs will be equal (because $\mathcal{T}_{1j}-\mathcal{T}_{2j} =0$ by \eqref{EQ:t1j:minus:t2j:c1:c2}). Hence it randomizes equally between the two BSs. First assume that $X_{ij} \neq 0 \; \forall i \in \{1,2\}, j \in \{1,2\}$. Then: 
\begin{eqnarray} 
 \Delta \tilde{V}(s)&\hspace{-0.5cm}= &\hspace{-0.5cm}\lambda p_1\left(\frac{1}{c_2}\right)+ \frac{\lambda p_2}{2}\left(\frac{1}{c_2} \right)+ \frac{\lambda p_2}{2}\left(\frac{1}{c_1} \right)\nonumber\\
& \hspace{0.1cm}+&\hspace{-0.3cm}\frac{\mu X_{11} c_1}{X_{11}+X_{12}}\left(\frac{-1}{ c_1}\right)+\frac{\mu X_{12} c_2}{X_{11}+X_{12}}\left(\frac{-1}{ c_2}\right)\nonumber\\
& \hspace{0.1cm}+&\hspace{-0.3cm}\frac{\mu X_{21} c_1}{X_{21}+X_{22}}\left(\frac{-1}{ c_1}\right)+\frac{\mu X_{22}c_2}{X_{21}+X_{22}}\left(\frac{-1}{ c_2}\right).
 \end{eqnarray} 
So $\Delta \tilde{V}(s)= \frac{5\lambda}{8}-2\mu < \frac{5}{8}(\lambda - \frac{4\mu}{3})< -\epsilon.$ 

Next, if $X_{11}=X_{12}=0$ and $X_{21}+X_{22} \neq 0$, then similar to the above derivation, it can be shown that:
 $\Delta \tilde{V}(s)= \frac{5\lambda}{8}-\mu < \frac{5}{8}(\lambda - \frac{4\mu}{3})< -\epsilon.$

If $X_{21}=X_{22}=0$ and $X_{11}+X_{12} \neq 0$, then similar to the above derivation:
 $\Delta \tilde{V}(s)= \frac{5\lambda}{8}-\mu < \frac{5}{8}(\lambda - \frac{4\mu}{3})< -\epsilon.$

\textbf{Case 2: $S_2 > S_1$}\\
If a new user $j$ with rate vector $(c_1,c_2)$ or $(c_2,c_1)$ arrives, it  chooses BS 1 (because $\mathcal{T}_{1j}-\mathcal{T}_{2j} >0$). First, assume that $X_{ij} \neq 0 \; \forall i \in \{1,2\}, j \in \{1,2\}$. Then:
\begin{eqnarray}
\Delta \tilde{V}(s)&\hspace{-0.5cm}= &\hspace{-0.5cm}\lambda p_1\left(\frac{1}{ c_2}\right)+ \lambda p_2\left(\frac{1}{ c_1} \right)+\frac{\mu X_{11} c_1}{X_{11}+X_{12}}\left(\frac{-1}{ c_1}\right)\nonumber\\
& \hspace{0.5cm}+&\frac{\mu X_{12} c_2}{X_{11}+X_{12}}\left(\frac{-1}{ c_2}\right)+\frac{\mu X_{21} c_1}{X_{21}+X_{22}}\left(\frac{-1}{ c_1}\right)\nonumber\\
& \hspace{0.5cm}+&\frac{\mu X_{22} c_2}{X_{21}+X_{22}}\left(\frac{-1}{ c_2}\right).
\end{eqnarray}

So $\Delta \tilde{V}(s)= \frac{3\lambda}{4}-2\mu < \frac{3}{4}(\lambda - \frac{4\mu}{3})< -\epsilon.$ 

Next, if $X_{11}=X_{12}=0$ and $X_{21}+X_{22} \neq 0$, then similar to the above derivation:\\
$\Delta \tilde{V}(s)= \frac{3\lambda}{4}-\mu = \frac{3}{4}(\lambda - \frac{4\mu}{3})< -\epsilon.$

If $X_{21}=X_{22}=0$ and $X_{11}+X_{12} \neq 0$, then similar to the above derivation:
$\Delta \tilde{V}(s)= \frac{3\lambda}{4}-\mu = \frac{3}{4}(\lambda - \frac{4\mu}{3})< -\epsilon.$ 

\textbf{Case 3: $S_1 > S_2$}, If a new user $j$ with rate vector $(c_1,c_2)$ arrives, it chooses BS 2 (because $\mathcal{T}_{1j}-\mathcal{T}_{2j} <0$).\\
\textbf{Case 3(a)}: $4S_2 > S_1$ 

If a new user $j$ with rate vector $(c_2,c_1)$ arrives, then it chooses BS 1. First, assume that $X_{ij} \neq 0 \; \forall i \in \{1,2\}, j \in \{1,2\}$. Then:
\begin{eqnarray}
\Delta \tilde{V}(s)&\hspace{-0.5cm}= &\hspace{-0.5cm}\lambda p_1\left(\frac{1}{ c_2}\right)+ \lambda p_2\left(\frac{1}{ c_2} \right)+\frac{\mu X_{11} c_1}{X_{11}+X_{12}}\left(\frac{-1}{ c_1}\right)\nonumber\\
& \hspace{0.5cm}+&\frac{\mu X_{12} c_2}{X_{11}+X_{12}}\left(\frac{-1}{ c_2}\right)+\frac{\mu X_{21} c_1}{X_{21}+X_{22}}\left(\frac{-1}{ c_1}\right)\nonumber\\
& \hspace{0.5cm}+&\frac{\mu X_{22} c_2}{X_{21}+X_{22}}\left(\frac{-1}{ c_2}\right).
\end{eqnarray}

So $\Delta \tilde{V}(s)= \frac{\lambda}{2}-2\mu < \frac{1}{2}(\lambda - \frac{4\mu}{3})= -\epsilon.$

Next, if $X_{11}=X_{12}=0$ and $X_{21}+X_{22} \neq 0$, then:
 $\Delta \tilde{V}(s)= \frac{\lambda}{2}-\mu < \frac{1}{2}(\lambda - \frac{4\mu}{3})= -\epsilon.$ 

If $X_{21}=X_{22}=0$ and $X_{11}+X_{12} \neq 0$, then:
 $\Delta \tilde{V}(s)= \frac{\lambda}{2}-\mu < \frac{1}{2}(\lambda - \frac{4\mu}{3})= -\epsilon.$

\textbf{Case 3(b)}: $4S_2 < S_1$ 

 If a new user $j$ with rate vector $(c_2,c_1)$ arrives, then it chooses BS 2. First, assume that $X_{ij} \neq 0 \; \forall i \in \{1,2\}, j \in \{1,2\}$. Then:
\begin{eqnarray}
\Delta \tilde{V}(s)&\hspace{-0.5cm}= &\hspace{-0.5cm}\lambda p_1\left(\frac{1}{ c_1}\right)+ \lambda p_2\left(\frac{1}{ c_2} \right)+\frac{\mu X_{11} c_1}{X_{11}+X_{12}}\left(\frac{-1}{ c_1}\right)\nonumber\\
& \hspace{0.5cm}+&\frac{\mu X_{12} c_2}{X_{11}+X_{12}}\left(\frac{-1}{ c_2}\right)+\frac{\mu X_{21} c_1}{X_{21}+X_{22}}\left(\frac{-1}{ c_1}\right)\nonumber\\
& \hspace{0.5cm}+&\frac{\mu X_{22} c_2}{X_{21}+X_{22}}\left(\frac{-1}{ c_2}\right).
\end{eqnarray}

So $\Delta \tilde{V}(s)= \frac{3\lambda}{4}-2\mu < \frac{3}{4}(\lambda - \frac{4\mu}{3})< -\epsilon.$ 
 
If $X_{11}=X_{12}=0$ and $X_{21}+X_{22} \neq 0$, then:
 $\Delta \tilde{V}(s)= \frac{3\lambda}{4}-\mu = \frac{3}{4}(\lambda - \frac{4\mu}{3})< -\epsilon.$ 

If $X_{21}=X_{22}=0$ and $X_{11}+X_{12} \neq 0$, then
 $\Delta \tilde{V}(s)= \frac{3\lambda}{4}-\mu = \frac{3}{4}(\lambda - \frac{4\mu}{3})< -\epsilon.$

\textbf{Case 3(c)}:  $4S_2 =S_1$ 

If  a new user $j$ with rate vector $(c_2,c_1)$ arrives, then it randomizes equally between the two BSs. If $X_{ij} \neq 0 \; \forall i \in \{1,2\}, j \in \{1,2\}$, then:
\begin{eqnarray} 
 \Delta \tilde{V}(s)&\hspace{-0.5cm}= &\hspace{-0.5cm}\lambda p_2\left(\frac{1}{c_2}\right)+ \frac{\lambda p_1}{2}\left(\frac{1}{c_2} \right)+ \frac{\lambda p_1}{2}\left(\frac{1}{c_1} \right)\nonumber\\
& \hspace{0.1cm}+&\hspace{-0.3cm}\frac{\mu X_{11} c_1}{X_{11}+X_{12}}\left(\frac{-1}{ c_1}\right)+\frac{\mu X_{12} c_2}{X_{11}+X_{12}}\left(\frac{-1}{ c_2}\right)\nonumber\\
& \hspace{0.1cm}+&\hspace{-0.3cm}\frac{\mu X_{21} c_1}{X_{21}+X_{22}}\left(\frac{-1}{ c_1}\right)+\frac{\mu X_{22}c_2}{X_{21}+X_{22}}\left(\frac{-1}{ c_2}\right).
 \end{eqnarray} 

So: 
$\Delta \tilde{V}(s)= \frac{5\lambda}{8}-2\mu < \frac{5}{8}(\lambda - \frac{4\mu}{3})< -\epsilon.$ 
 
If $X_{11}=X_{12}=0$ and $X_{21}+X_{22} \neq 0$, then:
$\Delta \tilde{V}(s)= \frac{5\lambda}{8}-\mu < \frac{5}{8}(\lambda - \frac{4\mu}{3})< -\epsilon.$

If $X_{21}=X_{22}=0$ and $X_{11}+X_{12} \neq 0$, then:
$\Delta \tilde{V}(s)= \frac{5\lambda}{8}-\mu < \frac{5}{8}(\lambda - \frac{4\mu}{3})< -\epsilon.$

Thus, in all the cases, $\Delta \tilde{V}(s) \leq -\epsilon$. Hence, condition \eqref{stable3} in Theorem~\ref{stable} is satisfied if $\lambda  < \frac{4\mu}{3}$. It is easy to check that the other conditions in Theorem~\ref{stable} are also satisfied. 

The result follows.
 \end{IEEEproof}

\begin{IEEEproof}[\quad Proof of Theorem~\ref{thm:heuristic}] 
The proof of part (a) is similar to that of Theorem 6 in~\cite{gsk} and is omitted for brevity.

{\bf Proof of part (b):}

We prove the result using Theorem~\ref{stable}. Since $p_1 = 1$, $X_{11} = X_{22} = 0$. 

First, note that $X_{21}$ can never exceed $4$. This can be shown as follows. Suppose $X_{21} = 4$. A user arrives with rate vector $(c_2,c_1)$. The functions $g(\cdot)$ in (\ref{mixedeq})
for the two BSs are given by: 
\[ g(1) = \mathcal{T}_{1j} + 0.4, \, \, g(2) = \mathcal{T}_{2j} + 0.2 \]
Since $X_{21} = 4$, $\mathcal{T}_{2j}= \frac{m_2 c_1}{X_{21}+1} = 0.2$. From this and the fact that $\mathcal{T}_{1j} > 0$, it follows that $g(1) > g(2)$. Hence 
the user chooses BS 1 and $X_{21}$ will stay at 4. 

Next, note that the Mixed policy induces a 
CTMC with state $s = (X_{12} \, \, X_{21})$.
Consider the Lyapunov function:
\begin{equation}
V(s) = (X_{12} + \gamma X_{21})^2 
\end{equation}
where $\gamma = \sqrt{3} - 1$. Let $\Delta\tilde{V} (s) =\Delta V(s) \nu(s)$, where $\nu(s)$ is the total transition rate out of state $s$. 
Let $\epsilon > 0$ be any positive number. 
Let 
$X_{\epsilon}$ be a large number. The particular conditions that 
 $X_{\epsilon}$ must satisfy will be stated later. 
Consider the set of states $A = \{s: X_{12} > X_{\epsilon}\}$.
We will show that  $\Delta \tilde{V} (s) \le - \epsilon$ for all states $s \in A$. 
The complement of $A$ is the finite set $S_0$ in Theorem~\ref{stable}. 
Let $X_{\epsilon}$ be large enough so that if $X_{21} \le 3$ and an arrival occurs, BS 2 is selected. If $X_{21} = 4$ and an arrival occurs, BS 1 is 
selected. We must have one of the following three cases for state $s$:
\begin{enumerate}[(I)]
\item
$X_{21} = 0$:
\begin{eqnarray*}
 \Delta \tilde{V} (s) & = & \lambda [ (X_{12} + \gamma)^2 - X_{12}^2] \\
& & + \mu m_1 c_2 [(X_{12} - 1)^2 - X_{12}^2] \\
& = & 2 X_{12} \gamma ( \lambda - 2(\sqrt{3}+1) \mu) + \lambda \gamma^2 + \mu m_1 c_2.
\end{eqnarray*}
The first term on the RHS is negative for $\lambda < (\sqrt{3} + 3) \mu$. The second and third terms are constants. Hence, $\Delta \tilde{V} (s) < - \epsilon$
for $X_\epsilon$, and hence $X_{12}$, large enough.  
\item
\label{hstable:nu:case2}
$1 \le X_{21} \le 3$: $\; \Delta \tilde{V} (s)= $ 
\begin{eqnarray*}
\lambda [(X_{12} + \gamma X_{21} + \gamma)^2 - (X_{12} + \gamma X_{21})^2] \nonumber\\
&  \hspace{-6.3cm}+ \mu m_1 c_2 [(X_{12} + \gamma X_{21} - 1)^2 - (X_{12} + \gamma X_{21})^2]  \nonumber\\
&  \hspace{-6.3cm}+ \mu m_2 c_1 [(X_{12} + \gamma X_{21} - \gamma)^2 - (X_{12} + \gamma X_{21})^2].  \nonumber
\end{eqnarray*}
After simplification, $\; \Delta \tilde{V} (s)= $
\begin{eqnarray*}
 \hspace{0.4cm}2 \gamma X_{12} [\lambda - (2\sqrt{3}+3)\mu] + \lambda [2 \gamma^2 X_{21} + \gamma^2] \nonumber\\
& \hspace{-6.7cm} + \mu m_1 c_2 [-2 \gamma X_{21} + 1] + \mu m_2 c_1  [- 2 \gamma^2 X_{21} + \gamma^2]. 
\end{eqnarray*}
The first term on the RHS is negative for $\lambda < (\sqrt{3} + 3) \mu$ and the remaining terms  are bounded. Hence, $\Delta \tilde{V} (s) < - \epsilon$
for $X_\epsilon$, and hence $X_{12}$, large enough.  
\item
$X_{21} = 4$: $\; \Delta \tilde{V} (s)=$
\begin{eqnarray*} 
\hspace{0.3cm}\lambda \left[2(X_{12}+\gamma X_{21})+1 \right]-\mu m_1 c_2\left[2(X_{12}+\gamma X_{21})-1\right]\nonumber\\
&\hspace{-12cm}-\mu m_2 c_1\gamma\left[2(X_{12}+\gamma X_{21})-\gamma\right].
\end{eqnarray*}
After simplification, $\; \Delta \tilde{V} (s)= $
\begin{equation}
\hspace{0.3cm}2(X_{12}+\gamma X_{21})\left[\lambda -(3+\sqrt{3})\mu \right]+\lambda+\mu c_1 \gamma^2+2\mu c_2.
\end{equation}
The first term on the RHS is negative for $\lambda < (\sqrt{3} + 3) \mu$ and the remaining terms  are constant. Hence, $\Delta \tilde{V} (s) < - \epsilon$
for $X_\epsilon$, and hence $X_{12}$, large enough. 

Thus, in all the cases, $\Delta \tilde{V}(s) \leq -\epsilon$. Hence, condition \eqref{stable3} in Theorem~\ref{stable} is satisfied. It is easy to check that the other conditions in Theorem~\ref{stable} are also satisfied. 

The result follows.
\end{enumerate}

\end{IEEEproof}

\begin{IEEEproof}[\quad Proof of Theorem~\ref{cap_gen}] 
We first prove the ``only if" part, which states that for every association policy the system is unstable for $\lambda \geq (B-K+K c_2) \mu$. We will use Theorem~\ref{unstable} to prove the ``only if" part. Consider the Lyapunov function $V(i)=c_2\sum_{i=1}^{K}\Big(\frac{X_{11}}{c_1}+\frac{X_{12}}{c_2}\Big)+c_1 \sum_{i=K+1}^{B} \Big(\frac{X_{i1}}{c_1}+\frac{X_{i2}}{c_2}\Big)$ and and let the finite set $\mathcal{S}_0=\begin{Bmatrix}
\begin{pmatrix}
0   &   0   \\
\vdots & \vdots\\
0     & 0
\end{pmatrix}
\end{Bmatrix}$.

First assume that $X_{ir}>0, \forall i\in \mathcal{B},r\in\{1,2\}$. The case when some of them are zero is considered separately. We want to show that:
\begin{equation}
\Delta V(m)\ge 0, \quad \forall m\notin \mathcal{S}_0 
\end{equation} 
Let $\Delta \Tilde{V}(m)=\Delta V(m) \nu(m)$,
where $\nu(m)$ is the total transition rate out of state $m$ in (\ref{Eq8}). 
Note that 
\begin{equation}
\Delta \Tilde{V}(m)\geq 0 \Leftrightarrow \Delta V(m)\ge 0.
\end{equation}
Let the current state be $m$. The rate vector of an arriving user is one of $\mathcal{R}_i, i \in \{1,\cdots,K \}$ with equal probability. Now, consider the most general association policy in which an arriving user with rate vector $\mathcal{R}_i$ associates with BS $j$ with probability $q_{ij}$. Then, $\Delta \Tilde{V}(m)$
\begin{align}
&=\frac{\lambda}{K}  \Big[q_{11}c_2 \Big(\frac{1}{c_2}\Big)+q_{12}c_2 \Big(\frac{1}{c_1}\Big)+\cdots +q_{1K}c_2\Big(\frac{1}{c_1}\Big)  \nonumber\\
&\hspace{.5cm}+q_{1{(K+1)}}c_1 \Big(\frac{1}{c_1}\Big)+\cdots +q_{1B}c_1\Big(\frac{1}{c_1}\Big) \Big] \nonumber \\
&\hspace{.5cm}+ \frac{\lambda}{K} \Big[q_{21}c_2\Big(\frac{1}{c_1}\Big)+q_{22}c_2 \Big(\frac{1}{c_2}\Big)+\cdots +q_{2B}c_2\Big(\frac{1}{c_1}\Big)  \nonumber\\
&\hspace{.5cm}+q_{2{(K+1)}}c_1 \Big(\frac{1}{c_1}\Big)+\cdots +q_{2B}c_1\Big(\frac{1}{c_1}\Big) \Big] \nonumber \\
&\hspace{.5cm}+ \cdots \nonumber\\
&\hspace{.5cm}+ \frac{\lambda}{K} \Big[q_{K1}c_2\Big(\frac{1}{c_1}\Big)+q_{K2}c_2\Big(\frac{1}{c_1}\Big)+\cdots +q_{KK}c_2\Big(\frac{1}{c_2}\Big)  \nonumber\\
&\hspace{.5cm}+q_{K{(K+1)}}c_1 \Big(\frac{1}{c_1}\Big)+\cdots +q_{KB}c_1\Big(\frac{1}{c_1}\Big) \Big] \nonumber \\
&\hspace{.5cm}- \mu \{Kc_2+(B-K)c_1\} \label{Eq34} \\
&=\frac{\lambda}{K}\Big\{ \Big(\sum_{i=1}^{K}q_{ii}\Big)+\Big(\sum_{i=1}^{K}\sum_{j=K+1}^{B}q_{ij}\Big)+\frac{c_2}{c_1}\Big(\sum_{i=1}^{K}\sum_{j=1,j\ne i}^{K}q_{ij}\Big) \Big\} \nonumber\\
&\hspace{.5cm}- \mu \{Kc_2+(B-K)c_1\} \nonumber \\
&> \lambda-\mu(B-K+Kc_2) \ge 0 \nonumber
\end{align}
The first inequality follows since $\sum_{j \in \mathcal{B}} q_{ij}=1 \; \forall i \in \{ 1,...,K\}$ and $c_2 > c_1$. The second inequality follows since $\lambda\ge (B-K+K c_2) \mu$.

Now, consider the case when some out of the $X_{ir}, i\in\mathcal{B}, r\in\{1,2\}$ are zero. If $X_i>0,$ but exactly one of $X_{i1}$ and $X_{i2}$ is zero  for some/ all $i\in\mathcal{B}$, then (\ref{Eq34}) remains the same. However, if $X_i=0$ for $i \in \{1,\cdots,K\}$ then the $-\mu c_2$ term corresponding to each $i$ for which $X_i=0$ drops out in (\ref{Eq34}) and if $X_i=0$ for $i\in\{K+1,\cdots ,B\}$ then $-\mu c_1$ term corresponding to each $i$ for which $X_i=0$ drops out in (\ref{Eq34}). Hence $\Delta \Tilde{V}(m)\ge 0$ continues to hold.

Thus,~\eqref{unstable1} in Theorem~\ref{unstable} holds. It is easy to check that all the other conditions in Theorem~\ref{unstable} hold. Hence, the system is unstable if $\lambda \geq (B-K+K c_2) \mu$.

Now, we prove the ``if" part, which states that there exists an association policy such that the system is stable for $\lambda < (B-K+K c_2) \mu$. Suppose each arriving user that has rate vector $R_i, \;i\in \{1,\cdots,K\}$ associates with BS $i$ at rate $c_2$ with probability $(1-q_i)$ and with a BS selected uniformly at random from the set $\{K+1,\cdots,B\}$ at rate $c_1$ with probability $q_i,$ where $q_i=\frac{B-K}{B-K+Kc_2}$. The fraction of users associate with BS $i\in \{K+1,\cdots,B\}$ at rate $c_1$ is $\frac{1}{B-K+Kc_2}$. Hence, a CTMC is induced at BS $i\in \{1,\cdots,K\}$ with state $s=(X_{i 2})$ and at BS $i \in \{K+1, \cdots, B\}$ with state $s=(X_{i1})$. Consider the Lyapunov function $V(s)=X_{i2}$ for BS $i\in \{1,\cdots,K\}$  and $V(s)=X_{i1}$ for $i\in \{K+1,\cdots,B\}$ and finite state $\mathcal{S}_0=\{0\}$, $\forall i \in \mathcal{B}$.

At BS $i \in \{1,\cdots,K\}$ and for $s\notin \mathcal{S}_0$
\begin{align*}
\Delta \Tilde{V}(s)&=\lambda p_i (1-q_i)-\mu c_2 \\
&=\frac{\lambda c_2}{B-K+K c_2}-\mu c_2 
\end{align*}
Therefore, for $\lambda <(B-K+K c_2)\mu$, $\Delta \Tilde {V}(s) \le -\epsilon$, where $\epsilon= \mu c_2 - \frac{\lambda c_2}{B-K+K c_2}>0$.  

At BS $i \in \{K+1,\cdots,B\}$ and for $s\notin \mathcal{S}_0$
\begin{align*}
\Delta \Tilde{V}(s)&=\frac{\lambda}{B-K+K c_2}-\mu c_1 \\
&=\frac{1}{B-K+K c_2}\{\lambda-(B-K+K c_2)\mu c_1 \} 
\end{align*}
Therefore, for $\lambda <(B-K+K c_2)\mu$, $\Delta \Tilde {V}(s) \le -\epsilon$, where $\epsilon= \mu c_2 - \frac{\lambda c_2}{B-K+K c_2}>0$.
Hence,~\eqref{stable3} in Theorem~\ref{stable} holds. It is easy to check that other conditions in Theorem~\ref{stable} also hold. Hence, the system is stable if $\lambda <(B-K+K c_2)\mu$.

The result follows.
\end{IEEEproof}

\begin{IEEEproof}[Proof of Theorem~\ref{snr2}:]

Under the SNR based user association policy, an arriving user having rate vector $\mathcal{R}_i$ associates with BS $i$ at rate $c_2$. By the Poisson splitting property~\cite{wolfe}, there are independent Poisson arrivals at rate $\lambda p_i$ at BS $i\; \forall i\in\mathcal{B}$. 

We first prove the ``if" part. At any BS $i \in \mathcal{B}$, a CTMC is induced with state $s=(X_{i2})$, because $X_{i1}=0$. We will use Theorem~\ref{stable} to prove stability. Consider the Lyapunov function $V(s)=X_{i2}$ and $\mathcal{S}_0=\{s: X_{i2}\le m_i\}$. Let $\Delta \Tilde{V}(s)=\Delta V(s) \nu (s)$, where $\nu (s)=\lambda p_i + \mu c_2 \min (m_i, X_{i2})$ is the total rate of transitions out of state $s$. It is easy to see that $\nu (s)$ is upper and lower bounded by positive real constants. Hence, $\Delta \Tilde{V}(s)\le -\epsilon$ for some $\epsilon> 0$ and all $s \notin \mathcal{S}_0$ iff $\Delta V (s)\le - \epsilon'$ for some $\epsilon'> 0$ and all $s \notin \mathcal{S}_0$. For an arbitrary BS $i$ and $s \notin \mathcal{S}_0$,
\begin{equation}
\Delta \Tilde {V}(s)=\lambda p_i -\mu c_2 m_i. 
\end{equation}

Therefore, for $\lambda < \mu c_2 \frac{m_i}{p_i}$, $\Delta \Tilde {V}(s) \le -\epsilon$, where $\epsilon=\mu c_2 m_i -\lambda p_i >0$. The system is stable if it is stable at all the BSs in $\mathcal{B}$. Hence, the system is stable if $\lambda < \mu c_2 \min_{i\in \mathcal{B}} (\frac{m_i}{p_i})$.  

Using the same Lyapunov functions as above,  and Theorem~\ref{unstable}, it can be shown that the system is unstable if $\lambda > \mu c_2 \min_{i\in \mathcal{B}} (\frac{m_i}{p_i})$. We omit the details for brevity.

The result follows.
\end{IEEEproof}


\begin{IEEEproof}[Proof of Theorem~\ref{load2}:]

\textbf{Proof of part (a):}

We prove part (a) using Theorem 2. Consider the Lyapunov function
 $V(s)=\sum_{i \in \mathcal{B}}\left(\frac{X_{i1}}{c_1}+\frac{X_{i2}}{c_2}\right)$ and finite set $\mathcal{S}_0=\begin{Bmatrix}
 	\begin{pmatrix}
 		0   &   0   \\
 		\vdots & \vdots\\
 		0     & 0
 	\end{pmatrix}
 \end{Bmatrix}$. 
 Define $\Delta \Tilde {V}(s)=\Delta V(s)\nu (s)$, where $\nu (s)$ is the total rate at which transitions out of state $s$ occur and is given by \eqref{Eq8}. It is easy to check that $\lambda \le \nu (s)\le \lambda+\mu c_2 \sum_{i=1}^{B}m_i$ for all $s\notin \mathcal{S}_0$. Hence, $\Delta \Tilde{V}(s)\geq 0 \Leftrightarrow \Delta V(s)\ge 0,\, \forall s \notin \mathcal{S}_0$. 
 
%

We define the following subsets of the set of BSs:\\
$\mathcal{I}=\{i\in \mathcal{B}|\, X_{i}= \min_{a \in \mathcal{B}} X_a\}$, $\mathcal{J}=\{i\in \mathcal{B}|\,m_i \le X_{i}\}$, $\mathcal{J}_1=\{i\in \mathcal{B}|X_{i1}>0, X_{i2}=0\}$, $\mathcal{J}_2=\{i\in \mathcal{B}|X_{i1}=0, X_{i2}>0\}$, $\mathcal{J}_3=\{i\in \mathcal{B}|X_{i1}>0, X_{i2}>0\}$  and $\mathcal{J}_4=\{i\in \mathcal{B}|X_{i1}=0, X_{i2}=0\}$. Then,
\vspace{-0.3cm}
\begin{align}
\vspace{2em}
&m_j \le X_{j1}, \qquad j \in {\mathcal{J}\cap \mathcal{J}_1}, \nonumber\\
&m_j \le X_{j2}, \qquad j \in {\mathcal{J}\cap \mathcal{J}_2}, \nonumber\\
&m_j \le X_{j}, \qquad j \in {\mathcal{J}\cap \mathcal{J}_3}, \label{Eq26}\\
&m_j > X_{j1}, \qquad j \in ({\mathcal{B} \backslash \mathcal{J}})\cap \mathcal{J}_1, \nonumber\\
&m_j > X_{j2}, \qquad j \in ({\mathcal{B} \backslash \mathcal{J}})\cap \mathcal{J}_2, \nonumber\\
&m_j > X_{j}, \qquad j \in ({\mathcal{B} \backslash \mathcal{J}})\cap \mathcal{J}_3. \nonumber
\end{align}

Now, first consider the case when $\mathcal{J}_4=\emptyset$, where $\emptyset$ represents the empty set. The case when $\mathcal{J}_4\neq \emptyset $ is dealt with later. Let $\mathcal{I}=\{i_1,...,i_K\}$ and $\mathcal{B} \backslash \mathcal{I}=\{j_1,...,j_{(B-K)}\}$; then $\Delta \Tilde{V}(s)$
\vspace{-0.2cm}
\begin{align}
&=\lambda \Big\{p_{i_1}\Big( \frac{1}{K}\frac{1}{c_2}+\frac{1}{K}\frac{1}{c_1}+\cdots+\frac{1}{K}\frac{1}{c_1}\Big) \nonumber\\
&\hspace{.5cm}+p_{i_2}\Big( \frac{1}{K}\frac{1}{c_1}+\frac{1}{K}\frac{1}{c_2}+\cdots+\frac{1}{K}\frac{1}{c_1}\Big)+\cdots \nonumber \\
&\hspace{.5cm}+p_{i_K} \Big( \frac{1}{K}\frac{1}{c_1}+\frac{1}{K}\frac{1}{c_1}+\cdots+\frac{1}{K}\frac{1}{c_2}\Big) \Big\} \nonumber \label{Eq27}\\
&\hspace{.5cm}+\lambda \Big\{p_{j_1}\Big( \frac{1}{K}\frac{1}{c_1}+\frac{1}{K}\frac{1}{c_1}+\cdots+\frac{1}{K}\frac{1}{c_1}\Big) \\
&\hspace{.5cm}+p_{j_2}\Big( \frac{1}{K}\frac{1}{c_1}+\frac{1}{K}\frac{1}{c_1}+\cdots+\frac{1}{K}\frac{1}{c_1}\Big)+\cdots \nonumber \\
&\hspace{.5cm}+p_{j_{(B-K)}} \Big( \frac{1}{K}\frac{1}{c_1}+\frac{1}{K}\frac{1}{c_1}+\cdots+\frac{1}{K}\frac{1}{c_1}\Big)  \Big\} \nonumber\\
&\hspace{.5cm}-\mu \Big\{\sum_{i=1}^{B}\left((\min (m_i, X_i) \frac{c_1 X_{i1}}{X_i}  \frac{1}{c_1} +\min (m_i, X_i) \frac{c_2 X_{i2}}{X_i}  \frac{1}{c_2}\right) \Big\}\nonumber
\end{align} 

 The first $K$ terms in the RHS of the above equation correspond to the arrival in the cells of the BSs in $\mathcal{I},$ whereas the next $(B-K)$ terms correspond to the arrival in the cells of the BSs in $\mathcal{B}-\mathcal{I}$.
 
After algebraic simplification, using~\eqref{Eq26} and substituting $c_1=1$ and $c_2=2$ into \eqref{Eq27}, we get $\Delta \Tilde{V}(s)$
\begin{align} 
&=\lambda \left( 1- \sum_{i\in \mathcal{I}}\frac{p_i}{2 K} \right)-\mu \sum_{j \in \mathcal{J}\cap(\mathcal{J}_1 \cup \mathcal{J}_2 \cup \mathcal{J}_3)} m_j  \nonumber\\
 &\hspace{.5cm}-\mu \left(\sum_{j \in ({\mathcal{B}- \mathcal{J}})\cap \mathcal{J}_1} X_{j1} +\sum_{j \in ({\mathcal{B}- \mathcal{J}})\cap \mathcal{J}_2} X_{j2} +\sum_{j \in ({\mathcal{B}- \mathcal{J}})\cap \mathcal{J}_3} X_{j} \right)
 \label{eq62} 
\end{align}
Now, consider the facts in \textit{Claim 1} and \textit{Claim 2}, which are as follows: 

\textbf{Claim 1:} $\sum_{j \in \mathcal{J}\cap(\mathcal{J}_1 \cup \mathcal{J}_2 \cup \mathcal{J}_3)} m_j 
+\sum_{j \in ({\mathcal{B}- \mathcal{J}})\cap \mathcal{J}_1} X_{j1} +\sum_{j \in ({\mathcal{B}- \mathcal{J}})\cap \mathcal{J}_2} X_{j2} +\sum_{j \in ({\mathcal{B}- \mathcal{J}})\cap \mathcal{J}_3} X_{j} \le \sum_{i=1}^{B} m_i$
\begin{IEEEproof}
This is a straightforward implication of \eqref{Eq26}.
\end{IEEEproof}
\textbf{Claim 2:} $\sum_{i\in \mathcal{I}}\frac{p_i}{2 K} \le \frac{\max (p_1, p_2,\cdots, p_B) }{2}.$

\begin{IEEEproof}
 \vspace{-0.5cm}
\begin{align}
& \hspace{0.2cm} p_i \le \max (p_1, p_2,\cdots , p_B), \, \forall i\in \mathcal{I}\nonumber \\
 \Rightarrow &\frac{p_i}{2 K} \le \frac{\max (p_1, p_2, \cdots, p_B)}{2 K}, \, \forall i\in \mathcal{I}\nonumber \\
 \Rightarrow &\sum_{i \in \mathcal{I}}\frac{p_i}{2 K} \le \sum_{i \in \mathcal{I}}\frac{\max (p_1, p_2, \cdots, p_B)}{2 K}\nonumber\\
 \Rightarrow &\sum_{i \in \mathcal{I}}\frac{p_i}{2 K} \le \frac{\max (p_1, p_2,\cdots, p_B) }{2} \nonumber
\end{align}
\end{IEEEproof}
Since 
 $\lambda \ge \frac{2 \mu \sum_{i=1}^{B}m_i}{2-\max(p_1, p_2, \cdots, p_B)},$ we get: 
\begin{equation}
\lambda\{ 1- \frac{\max (p_1, p_2,\cdots, p_B) }{2} \}-\mu \sum_{j \in \mathcal{B}} m_j\ge 0 
\label{lam}
 \end{equation}
 
Substituting from \textit{Claim 1}, \textit{Claim 2} and~\eqref{lam} into~\eqref{eq62}, we get $\Delta \Tilde{V}(s)\ge 0$.  Hence, \eqref{unstable1} in Theorem~\ref{unstable} holds. It is easy to show that the other conditions in Theorem~\ref{unstable} also hold. Hence, the system is unstable for the case $\mathcal{J}_4 =\emptyset$.
 
 Now, consider the case when $\mathcal{J}_4\ne \emptyset$; then $\mathcal{I}=\mathcal{J}_4$. Now, from \eqref{Eq27}, $\Delta \Tilde{V}(s)$
 \begin{align}
 &=\lambda \left( 1- \sum_{i\in \mathcal{I}}\frac{p_i}{2 K} \right)-\mu \sum_{j \in \mathcal{J}\cap(\mathcal{J}_1 \cup \mathcal{J}_2 \cup \mathcal{J}_3 )} m_j  \nonumber\\
 &\hspace{.2cm}-\mu \left(\sum_{j \in ({\mathcal{B}- \mathcal{J}})\cap \mathcal{J}_1} X_{j1} +\sum_{j \in ({\mathcal{B}- \mathcal{J}})\cap \mathcal{J}_2} X_{j2} +\sum_{j \in ({\mathcal{B}- \mathcal{J}})\cap \mathcal{J}_3} X_{j} \right)
 \label{eq64}
 \end{align}
Substituting from \textit{Claim 1}, \textit{Claim 2} and~\eqref{lam} into~\eqref{eq64}, we get $\Delta \Tilde{V}(s)\ge 0$. Hence, \eqref{unstable1} in Theorem~\ref{unstable} holds. It is easy to show that the other conditions in Theorem~\ref{unstable} also hold. Hence, the system is unstable for the case $\mathcal{J}_4 \neq \emptyset$.

The result follows.
 
\textbf{Proof of part (b):}
We now prove part (b) using Theorem 1. Consider the Lyapunov function $V(s)=\sum_{i=1}^{B}(X_{i1}+X_{i2})$ and let
$\Delta \Tilde {V}(s)=\Delta V(s)\nu (s)$. Let the finite
set $\mathcal{S}_0$ be as in part (a). It is easy to show that $\nu (s)$ is lower and
upper bounded by constants.
Hence, $\Delta \Tilde{V}(s)\le -\epsilon$ for some $\epsilon> 0$ and all $s \notin \mathcal{S}_0$ iff $\Delta V (s)\le - \epsilon'$ for some $\epsilon'> 0$ and all $s \notin \mathcal{S}_0$. Now for $s \notin \mathcal{S}_0$, we will show that $\Delta \Tilde{V}(s)\le -\epsilon$ for some $\epsilon >0$, when $\lambda < \mu $.

Consider a subset of BSs $\mathcal{I}=\{i\in \mathcal{B}|\, X_{i}= \min_{a \in \mathcal{B}} X_a\}$ and let $\mathcal{I}=\{i_1,...,i_K\}$ and $\mathcal{B} \backslash \mathcal{I}=\{j_1,...,j_{(B-K)}\}$.
First assume $X_i> 0$, $\forall i\in \mathcal{I}$. The case when $X_i=0$, $\forall i\in \mathcal{I}$ is dealt with later.  Then, for $s \notin \mathcal{S}_0$, $\Delta \Tilde{V}(s)$
\begin{align}
&=\lambda\Big\{ p_{i_1}\left(\frac{1}{K}+\cdots+\frac{1}{K}\right)+p_{i_2}\left(\frac{1}{K}+\cdots+\frac{1}{K}\right)+\cdots \nonumber \\
&\hspace{.5cm}+p_{i_K}\left(\frac{1}{K}+\cdots +\frac{1}{K}\right)\Big\}\\
&\hspace{.5cm}+\lambda\Big\{p_{j_1}\left(\frac{1}{K}+\cdots+\frac{1}{K}\right)+p_{j_2}\left(\frac{1}{K}+\cdots+\frac{1}{K}\right)+\cdots  \nonumber\\
&\hspace{.5cm}+p_{j_{(B-K)}}\left(\frac{1}{K}+\cdots+\frac{1}{K}\right) \Big\} \nonumber\\
&\hspace{.5cm}-\mu\Big\{\sum_{i=1}^{B}\min (m_i,X_i) \frac{c_1 X_{i1}+c_2 X_{i2}}{X_i}\Big \} \nonumber\\
&=\lambda-\mu\sum_{i=1}^{B} \frac{c_1 X_{i1}+c_2 X_{i2}}{X_i}  \nonumber\\
&\leq  \lambda-\mu \nonumber
\end{align}
The above inequality follows since $\sum_{i=1}^{B}\frac{c_1 X_{i1}+c_2 X_{i2}}{X_i} \ge 1$. Hence, for $\lambda <\mu$, $\Delta \Tilde {V}(s)<-\epsilon$, where $\epsilon = \mu - \lambda>0$. 

Now, consider the case when $X_i=0$, $\forall i\in \mathcal{I}$. The departure term corresponding to each of the BS $i\in \mathcal{I}$ in the above equation becomes zero. Hence, the above equation reduces to $\Delta \Tilde{V}(s)=\lambda-\mu\sum_{i \in \mathcal{B}-\mathcal{I}} \frac{c_1 X_{i1}+c_2 X_{i2}}{X_i}$. As $\sum_{i\in \mathcal{B}-\mathcal{I}}\frac{c_1 X_{i1}+c_2 X_{i2}}{X_i}\ge 1$, therefore, for $\lambda <\mu$, $\Delta \Tilde {V}(s)<-\epsilon$, where $\epsilon = \mu - \lambda>0$. 

Hence, \eqref{stable3} in Theorem~\ref{stable} holds. It is easy to check that the other conditions in Theorem~\ref{stable} also hold. Hence, the system is stable  if $\lambda < \mu$.

The result follows.
\end{IEEEproof}

\bibliographystyle{IEEEtran}
\bibliography{ref} 
 
%

\end{document}